\documentclass[10pt]{article}
\usepackage{multicol}
\usepackage{graphicx}
\usepackage{amsmath}
\usepackage[a4paper]{geometry}
\usepackage{hyperref}

\begin{document}

\begin{center}
{\Large \bf Multiparticle production and initial quasi-temperature
from proton induced carbon collisions at $p_{Lab}=31$ GeV/$c$}

\vskip.75cm

Pei-Pin Yang$^1${\footnote{E-mail: yangpeipin@qq.com}}, Mai-Ying
Duan$^{1,}${\footnote{E-mail: duanmaiying@sxu.edu.cn}}, Fu-Hu
Liu$^{1,}${\footnote{Corresponding author. E-mail:
fuhuliu@163.com; fuhuliu@sxu.edu.cn}}, Raghunath
Sahoo$^{2,}${\footnote{E-mail: Raghunath.Sahoo@cern.ch;
raghunath.phy@gmail.com}}

{\small\it $^1$Institute of Theoretical Physics and Department of
Physics and\\ State Key Laboratory of Quantum Optics and Quantum
Optics Devices, Shanxi University, Taiyuan, Shanxi 030006, China

$^2$Discipline of Physics, School of Basic Sciences, Indian
Institute of Technology Indore, Simrol, Indore 453552, India}
\end{center}

\vskip.5cm

{\bf Abstract:} The momentum spectra of charged pions ($\pi^+$ and
$\pi^-$) and kaons ($K^+$ and $K^-$), as well as protons ($p$),
produced in the beam protons induced collisions in a 90-cm-long
graphite target [proton-carbon ($p$-C) collisions] at the beam
momentum $p_{Lab}=31$ GeV/$c$ are studied in the framework of a
multisource thermal model by using Boltzmann distribution and
Monte Carlo method. The theoretical model results are
approximately in agreement with the experimental data measured by
the NA61/SHINE Collaboration. The related free parameters
(effective temperature, rapidity shifts, and fraction of
non-leading protons) and derived quantities (average transverse
momentum and initial quasi-temperature) under given experimental
conditions are obtained. It is shown that the considered free
parameters and derived quantities to be strongly dependent on
emission angle over a range from 0 to 380 mrad and weakly
dependent on longitudinal position (graphite target thickness)
over a range from 0 to 90 cm.
\\

\\


\vskip1.0cm

\begin{multicols}{2}

{\section{Introduction}}

High energy (relativistic) nucleus-nucleus (heavy ion) collisions
with nearly zero impact parameter (central collisions) are
believed to form Quark-Gluon Plasma (QGP) or quark
matter~\cite{1,2,3} in the laboratory. High energy nucleus-nucleus
collisions with large impact parameter are not expected to form
QGP due to low particle multiplicity yielding lower energy density
and temperature~\cite{4}. Small collision systems such as
proton-nucleus and proton-proton collisions at high energy,
produce usually low multiplicity, which are not expected to form
QGP, but are useful to study the multiparticle production
processes. However, a few of proton-nucleus and proton-proton
collisions at the LHC energies can produce high multiplicity due
to nearly zero ``impact parameter", which are possibly expected to
form QGP, where the concept ``impact parameter" or ``centrality"
used in nuclear collisions are used in proton-proton
collisions~\cite{5}. Degree of collectivity, long-range
correlations, strangeness enhancement etc., which are considered
as QGP-like signatures, are recently observed in these high
multiplicity events~\cite{6,7,8}.

Assuming nucleus-nucleus collisions as a mere superposition of
proton-proton collisions in the absence of any nuclear effects,
usually one considers proton-proton collisions as the baseline
measurements. On the other hand, proton-nucleus
collisions~\cite{8a,8b,8c,8d,8e} serve as studying the initial
state effects and making a bridge between
proton-proton~\cite{8f,8g,8h,8i,8j} to nucleus-nucleus
collisions~\cite{8k,8l,8m,8n,8o} while studying the multiparticle
production processes, though fewer particles are produced in
proton-nucleus collisions than in nucleus-nucleus collisions.

There are different types of models or theories being introduced
in the studies of high energy collisions~\cite{9,10}. Among these
models or theories, different versions of thermal and statistical
models~\cite{11,12,13,14} characterize some of the aspects of
high-energy nuclear collisions, while there are many other aspects
that are studied by other approaches. As a basic concept,
temperature is ineluctable to be used in analyses. In fact, not
only the ``temperature is surely one of the central concepts in
thermodynamics and statistical mechanics"~\cite{15}, but also it
is very important due to its extremely wide applications in
experimental measurements and theoretical studies in subatomic
physics, especially in high energy and nuclear physics.

In view of this importance, in this paper, we are interested in
the study of proton-nucleus collisions at high energy by using the
Boltzmann distribution and the Monte Carlo Method in the framework
of the multisource thermal model~\cite{16}. The theoretical model
results are compared with the experimental data of the beam
protons induced collisions in a 90-cm-long graphite target
[proton-carbon ($p$-C) collisions] at the beam momentum
$p_{Lab}=31$ GeV/$c$ measured by the NA61/SHINE
Collaboration~\cite{17} at the Super Proton Synchrotron (SPS), the
European Organisation for Nuclear Research or the European
Laboratory for Particle Physics (CERN).

The remainder of this paper is structured as follows. The
formalism and method are shortly described in Section 2. Results
and discussion are given in Section 3. In Section 4, we summarize
our main observations and conclusions.
\\

{\section{Formalism and method}}

According to the multisource thermal model~\cite{16}, it is
assumed that there are many local emission sources to be formed in
high energy collisions due to different excitation degrees,
rapidity shifts, reaction mechanisms, impact parameters (or
centralities). In the transverse plane, the local emission sources
with the same excitation degree form a (large) emission source. In
the rapidity space, the local emission sources with the same
rapidity shift form a (large) emission source. In the rest frame
of an emission source with a determined excitation degree, the
particles are assumed to be emitted isotropically.

In the rest frame of a given emission source, let $T$ denote the
temperature parameter. The particles with rest mass $m_0$ produced
in the rest frame of the emission source are assumed to have the
simplest Boltzmann distribution of momenta $p'$~\cite{18}. That is
\begin{align}
f_{p'}(p')=Cp'^2\exp\left(-\frac{\sqrt{p'^2+m_0^2}}{T} \right),
\end{align}
where $C$ is the normalization constant which is related to $T$.
As a probability density function, Eq. (1) is naturally normalized
to 1.

If we need to consider multiple sources, we can use a
superposition of different equations with different temperatures
and fractions. We have
\begin{align}
f_{p'}(p')=\sum_j
k_jC_jp'^2\exp\left(-\frac{\sqrt{p'^2+m_0^2}}{T_j} \right),
\end{align}
where $k_j$, $C_j$, and $T_j$ are the fraction, normalization
constant, and temperature for the $j$-th source or component. The
average temperature obtained from Eq. (2) is
$T=\sum_jk_jT_j/\sum_jk_j=\sum_jk_jT_j$ due to $\sum_jk_j=1$. The
derived parameter $T$ is the weighted average over various
components, but not the simple weighted sum.

It should be noted that $T$ or $T_j$ is not the ``real"
temperature of the emission source, but the effective temperature
due to the fact that the flow effect is not excluded in the
momentum spectrum. The ``real" temperature is generally smaller
than the effective temperature which contains the contribution of
collective radial flow effect. To disengage the thermal motion and
collective flow effect, one may use different methods such as the
blast-wave model~\cite{24,25} or any alternative
method~\cite{26,27}. As an example, we shall discuss shortly the
results of the blast-wave model in section 3.

The contribution of spin being small, is not included in Eq. (1).
The effect of chemical potential ($\mu$) is not included in Eq.
(1) as well, due to the fact that $\mu$ affects only the
normalization, but not the trend, of the spectrum if the spin
effect is neglected. Our previous work~\cite{27a} shows that the
spin effect together with $\mu \gg m_0$ or $\mu \ll m_0$ is so
small ($<1\%$) that we do not need to consider it in studying
momentum or transverse momentum spectra in high energy collisions.
Only the combination of spin and $\mu\approx m_0$ causes an
obvious effect, which is not the case in this paper.

In the Monte Carlo method~\cite{19,20}, let $R_{1,2,3,4}$ denote
random numbers distributed evenly in $[0,1]$. To obtain a concrete
value of $p'$ which satisfies Eq. (1) or one of the components in
Eq. (2), we can perform the solution of
\begin{align}
\int_0^{p'} f_{p'}(p'')dp'' < R_1 < \int_0^{p'+\delta p'}
f_{p'}(p'')dp'',
\end{align}
where $\delta p'$ denotes a small shift relative to $p'$.

Under the assumption of isotropic emission in the rest frame of
emission source, the emission angle $\theta'$ of the considered
particle has the probability density function:
\begin{align}
f_{\theta'}(\theta')=\frac{1}{2}\sin\theta'
\end{align}
which is a half sine distribution in $[0,\pi]$, and the azimuth
$\phi'$ obeys the probability density function
$f_{\phi'}(\phi')=1/(2\pi)$ which is an even distribution in
$[0,2\pi]$~\cite{20a}. In the Monte Carlo method, $\theta'$
satisfies
\begin{align}
\theta'=2\arcsin \left( \sqrt{R_2} \right)
\end{align}
which is the solution of $\int_0^{\theta'} (1/2)\sin\theta''
d\theta''=R_2$.

Considering $p'$ and $\theta'$ obtained from Eqs. (3) and (5), we
have the transverse momentum $p'_T$ to be
\begin{align}
p'_T=p'\sin\theta',
\end{align}
the longitudinal momentum $p_z'$ to be
\begin{align}
p'_z=p'\cos\theta',
\end{align}
the energy $E'$ to be
\begin{align}
E'=\sqrt{p'^2+m_0^2},
\end{align}
and the rapidity $y'$ to be
\begin{align}
y'\equiv\frac{1}{2}\ln\left(\frac{E'+p'_z}{E'-p'_z}\right).
\end{align}

In the center-of-mass reference frame or the laboratory reference
frame, the rapidity of the considered emission source is assumed
to be $y_x$ in the rapidity space. Then, the rapidity of the
considered particle in the center-of-mass or laboratory reference
frame is
\begin{align}
y=y'+y_x
\end{align}
due to the additivity of rapidity. Multiple emission sources are
assumed to distribute evenly in the rapidity range $[y_{\min},
y_{\max}]$, where $y_{\min}$ and $y_{\max}$ are the minimum and
maximum rapidity shifts of the multiple sources. In the Monte
Carlo method,
\begin{align}
y_x=(y_{\max}-y_{\min})R_3+y_{\min}.
\end{align}

In particular, comparing with small mass particles, protons
exhibit large effect of leading particles which are assumed to
distribute evenly in the rapidity range $[y_{L\min}, y_{L\max}]$,
where $y_{L\min}$ and $y_{L\max}$ are the minimum and maximum
rapidity shifts of the leading protons. We have
\begin{align}
y_x=(y_{L\max}-y_{L\min})R_4+y_{L\min}.
\end{align}
The fraction of the non-leading (leading) protons in total protons
is assumed to be $k$ ($1-k$). The effects of leading pions and
kaons are small and can be neglected in this paper.

In the center-of-mass or laboratory reference frame, the
transverse momentum $p_T$ is
\begin{align}
p_T=p'_T,
\end{align}
the longitudinal momentum $p_z$ is
\begin{align}
p_z=\sqrt{p_T^2+m_0^2}\sinh y,
\end{align}
the momentum $p$ is
\begin{align}
p=\sqrt{p_T^2+p_z^2},
\end{align}
and the emission angle $\theta$ is
\begin{align}
\theta=\arctan\left(\frac{p_T}{p_z}\right).
\end{align}

The whole calculation is performed by the Monte Carlo method,
though only random numbers are used for the numerical calculation.
To compare the theoretical model results with the experimental
momentum spectra in a given $\theta$ range, we analyze the
momentum distribution of particles which are in the given $\theta$
range. It should be noted that another experimental selection,
i.e. the longitudinal position $z$~\cite{17}, is not regarded as
the selected condition in the theoretical model work due to the
fact that $z$ is only a reflection of target thickness in a
90-cm-long graphite target. From $z=0$ to $z=90$ cm, the beam
momentum slightly decreases, which is neglected in this paper.
In the calculation using random numbers, the energy-momentum
conservation was demanded at each step. The results violating the
energy-momentum conservation are not considered for our discussions.

It should be noticed that the Boltzmann distribution, Eq. (1), can
be used to describe low momentum spectra in the source's rest
frame or low transverse momentum spectra after analytic
derivation~\cite{20a} or via the Monte Carlo method, Eqs. (3),
(5), and (6). In the case of considering high momentum spectra in
the source's rest frame or high transverse momentum spectra, one
may use possibly the multi-component Boltzmann distribution, Eq.
(2). This paper treats multiple sources moving directly in a
rapidity range, $[y_{\min}, y_{\max}]$ or $[y_{L\min},
y_{L\max}]$, which results in high momentum in laboratory
reference frame. However, in the rest frame of each source, the total
momentum and transverse momentum are small. As a consequence,
Eq. (1) is valid in all momentum range, after the transformation
from source's rest frame to laboratory reference frame.
\\

{\section{Results and discussion}}

\begin{figure*}[htbp]
\vskip-1.50cm
\begin{center}
\includegraphics[width=12.50cm]{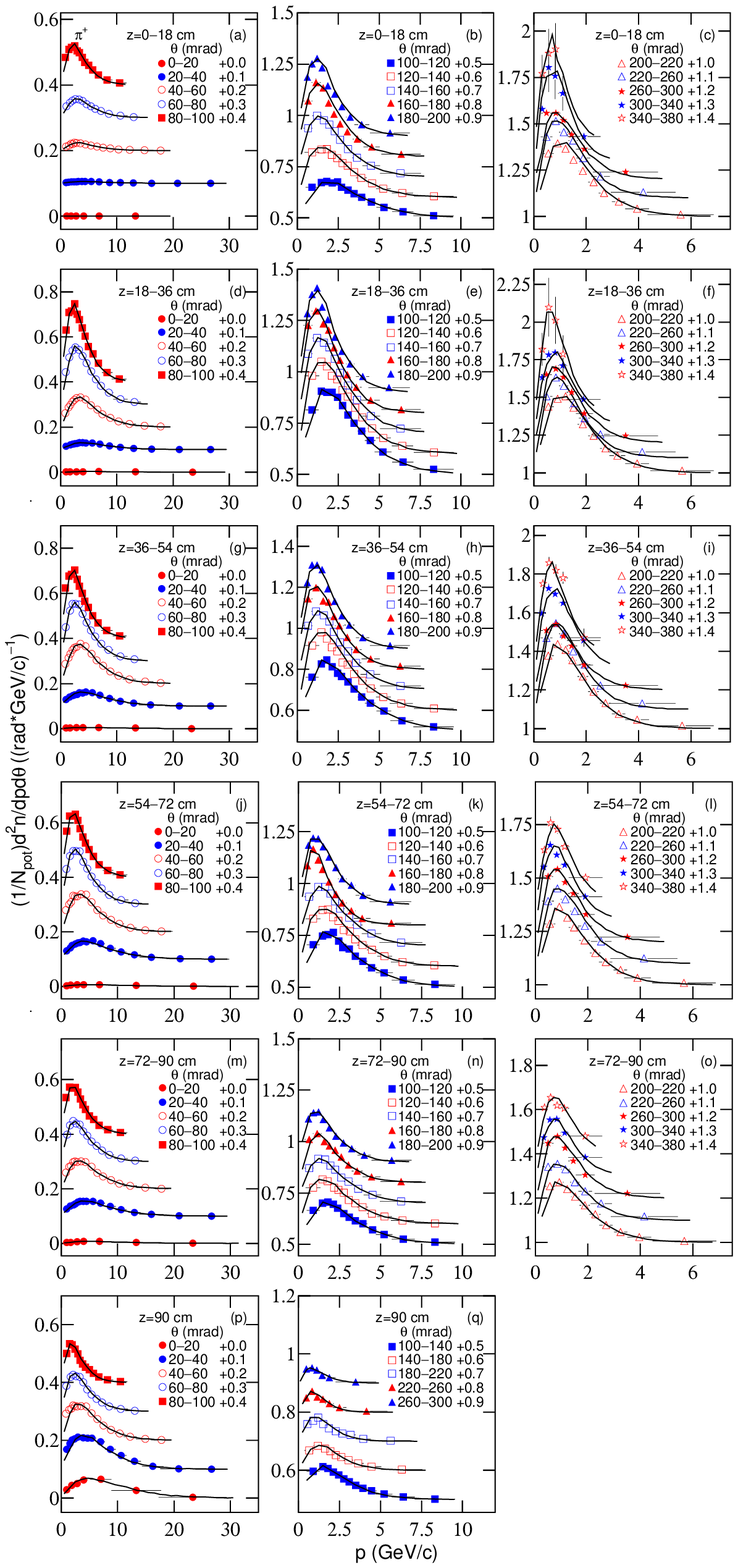}
\end{center}
\vskip-0.50cm {\small Fig. 1. Momentum spectra of $\pi^+$ produced
in $p$-C collisions at 31 GeV/$c$. Panels (a)--(c), (d)--(f),
(g)--(i), (j)--(l), (m)--(o), and (p)--(q) represent the spectra
for $z=0$--18, 18--36, 36--54, 54--72, 72--90, and 90 cm,
respectively. The symbols represent the experimental
data~\cite{17}. The curves are our results fitted by the
multisource thermal model due to Eq. (1) and Monte Carlo method.
To show clearly, different spectra are scaled by adding different
amounts marked in the panels.}
\end{figure*}

\begin{figure*}[htbp]
\vskip-1.50cm
\begin{center}
\includegraphics[width=12.50cm]{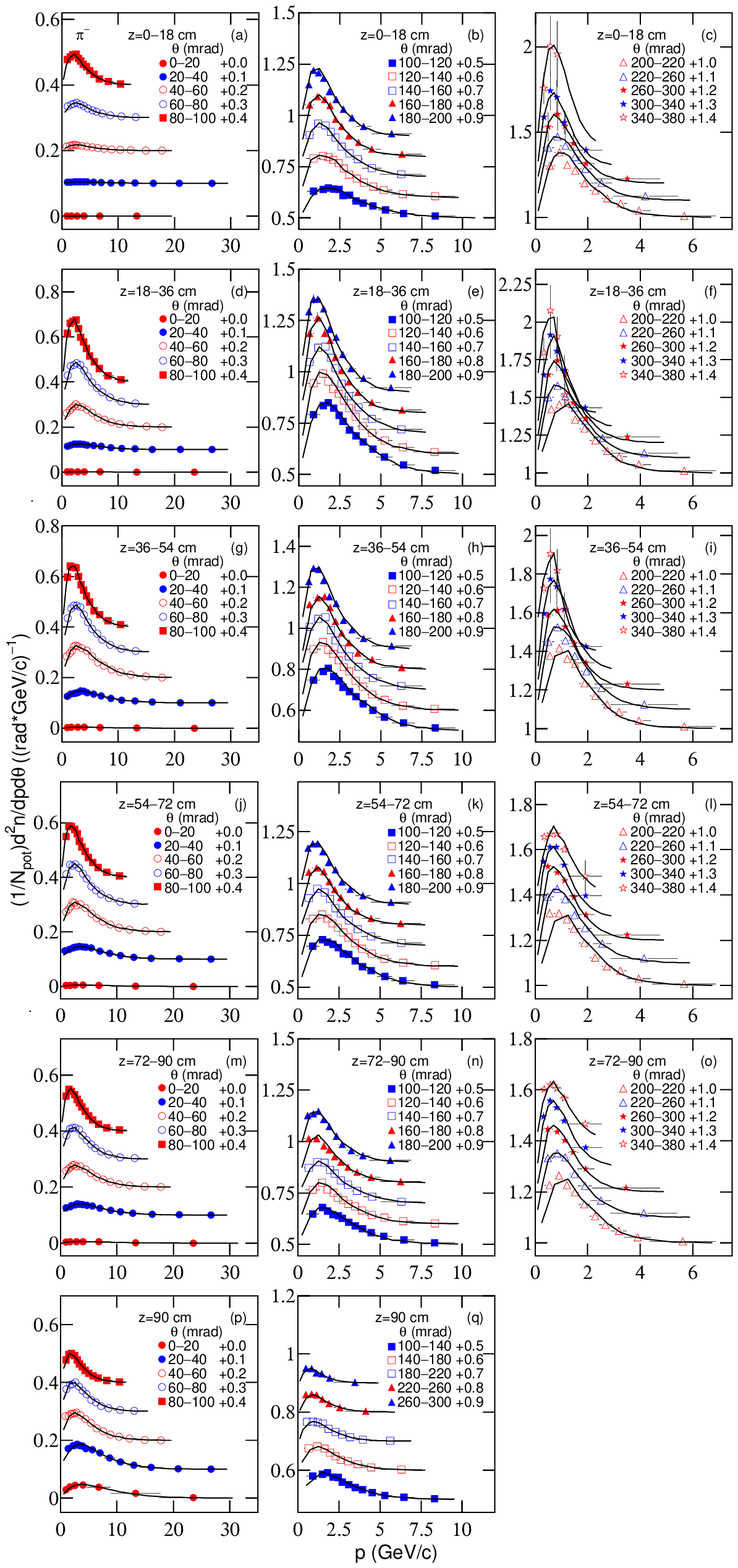}
\end{center}
\vskip-0.50cm {\small Fig. 2. Same as Fig. 1, but showing the
spectra of $\pi^-$.}
\end{figure*}

\begin{figure*}[!htb]
\vskip-1.0cm
\begin{center}
\includegraphics[width=12.50cm]{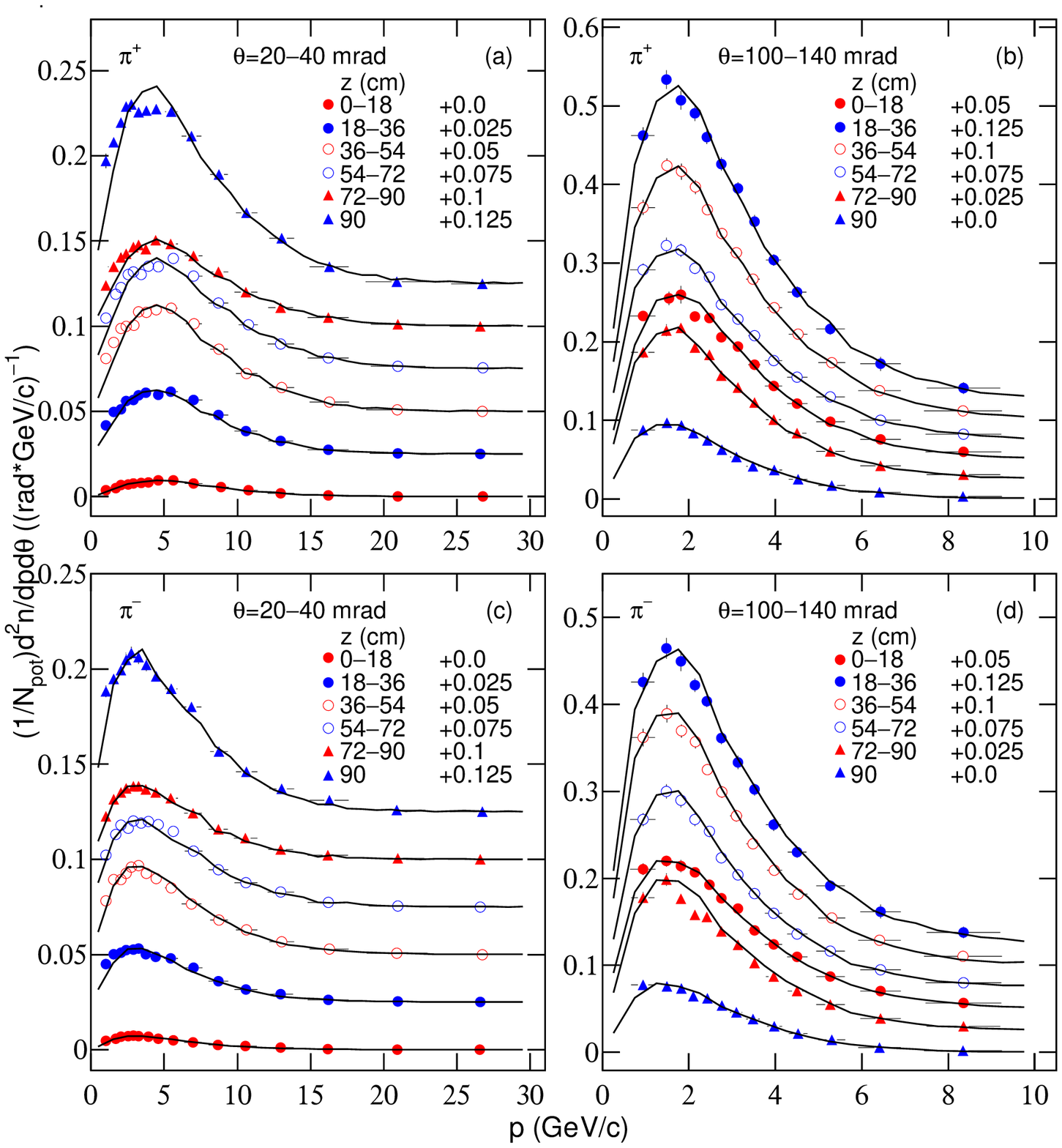}
\end{center}
\vskip-0.50cm {\small Fig. 3. Same as Fig. 1, but showing the
spectra of (a)--(b) $\pi^+$ and (c)--(d) $\pi^-$ in (a)--(c)
$\theta=$20--40 mrad and (b)--(d) $\theta=$100--140 mrad in six
$z$ ranges.}
\end{figure*}

Figures 1 and 2 present the momentum spectra, $(1/N_{pot})
d^2n/dpd\theta$, of charged pions ($\pi^+$ and $\pi^-$) produced
in $p$-C collisions at 31 GeV/$c$ in the laboratory reference
frame respectively, where $N_{pot}$ denotes the number of protons
on target and $n$ denotes the number of particles. Panels
(a)--(c), (d)--(f), (g)--(i), (j)--(l), (m)--(o), and (p)--(q)
represent the spectra for $z=0$--18, 18--36, 36--54, 54--72,
72--90, and 90 cm, respectively. For clarity, spectra in different
$\theta$ ranges are scaled by adding different numbers (marked in
the panels) are represented by different symbols, which are the
experimental data measured by the NA61/SHINE
Collaboration~\cite{17}. The curves are our results fitted by the
multisource thermal model using to Eq. (1) and Monte Carlo method.
The values of free parameters ($T$, $y_{\max}$ and $y_{\min}$),
normalization constant ($N_0$), $\chi^2$, and number of degree of
freedom (ndof) corresponding to the fits for the spectra of
$\pi^+$ and $\pi^-$ are listed in Tables A1 and A2 in the
appendix, respectively. In two cases, ndof in the fittings are
negative which appear in the tables with ``$-$" signs and the
corresponding curves are for eye guiding only. One can see that
the theoretical model results are approximately in agreement with
the NA61/SHINE experimental data of $\pi^+$ and $\pi^-$.

Figure 3 presents the momentum spectra of (a)--(b) $\pi^+$ and
(c)--(d) $\pi^-$ in (a)--(c) $\theta=20$--40 mrad and (b)--(d)
$\theta=100$--140 mrad in six $z$ ranges with different scaled
amounts shown in the panels. The symbols represent the
experimental data~\cite{17}. The curves are our results fitted by
the model. The values of $T$, $y_{\max}$, $y_{\min}$, $N_0$,
$\chi^2$, and ndof corresponding to the fits for the spectra of
$\pi^+$ and $\pi^-$ are listed in Table A3 in the appendix. One
can see again that the theoretical model results are approximately
in agreement with the experimental data of $\pi^+$ and $\pi^-$.

Similar to Figs. 1 and 2, Figs. 4 and 5 show the momentum spectra
of positively and negatively charged kaons ($K^+$ and $K^-$)
produced in $p$-C collisions at 31 GeV/$c$ respectively. Panels
(a), (b), (c), (d), (e), and (f) represent the spectra for
$z=0$--18, 18--36, 36--54, 54--72, 72--90, and 90 cm,
respectively. The values of $T$, $y_{\max}$, $y_{\min}$, $N_0$,
$\chi^2$, and ndof corresponding to the fits for the spectra of
$K^+$ and $K^-$ are listed in Tables A4 and A5 respectively in the
appendix. One can see that the theoretical model results are
approximately in agreement with the experimental data of $K^+$ and
$K^-$.

\begin{figure*}[htbp]
\begin{center}
\includegraphics[width=12.50cm]{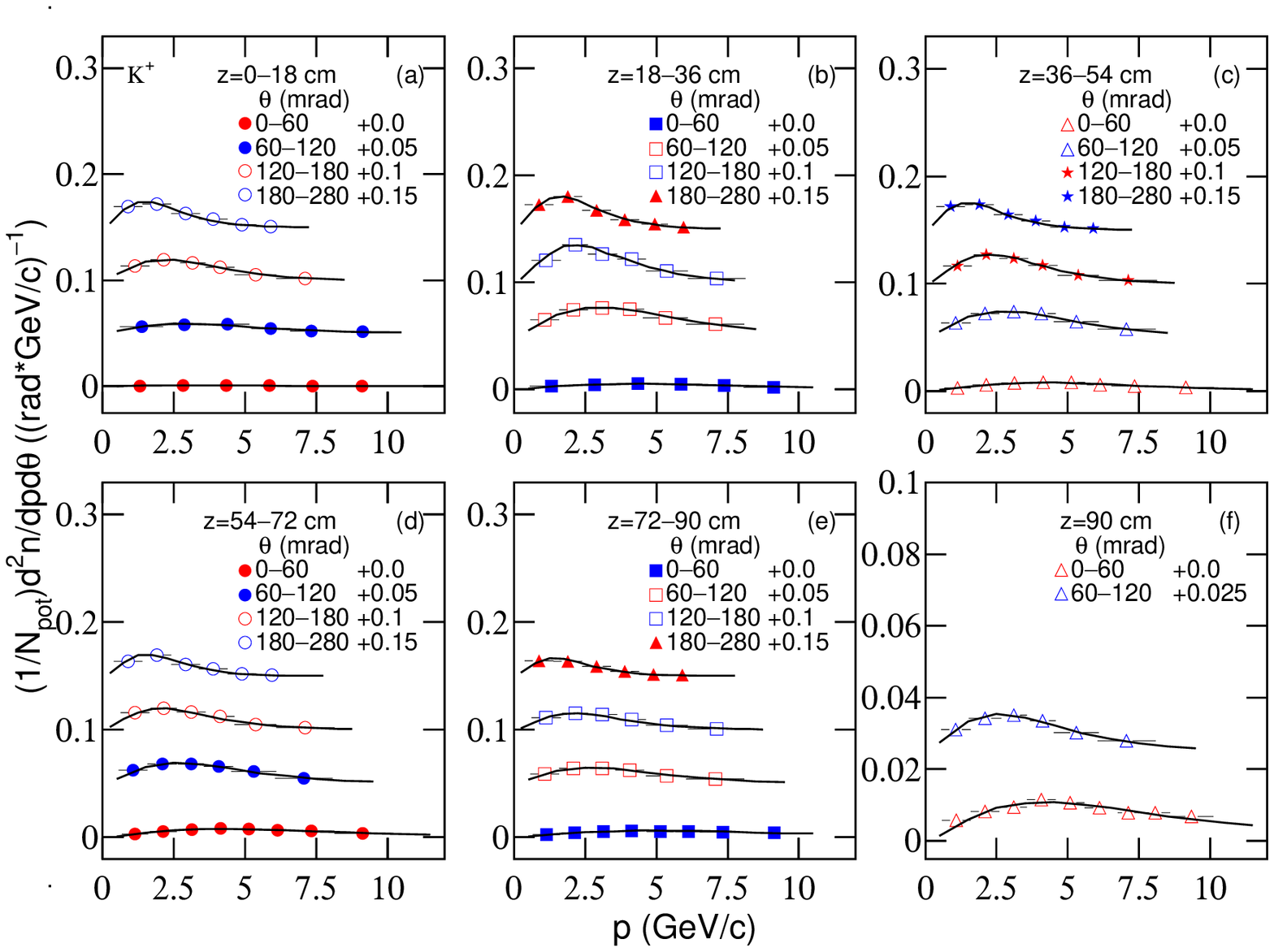}
\end{center}
{\small Fig. 4. Same as Fig. 1, but showing the spectra of $K^+$.
Panels (a), (b), (c), (d), (e), and (f) represent the spectra for
$z=0$--18, 18--36, 36--54, 54--72, 72--90, and 90 cm,
respectively.}
\end{figure*}

\begin{figure*}[!htb]
\begin{center}
\includegraphics[width=12.50cm]{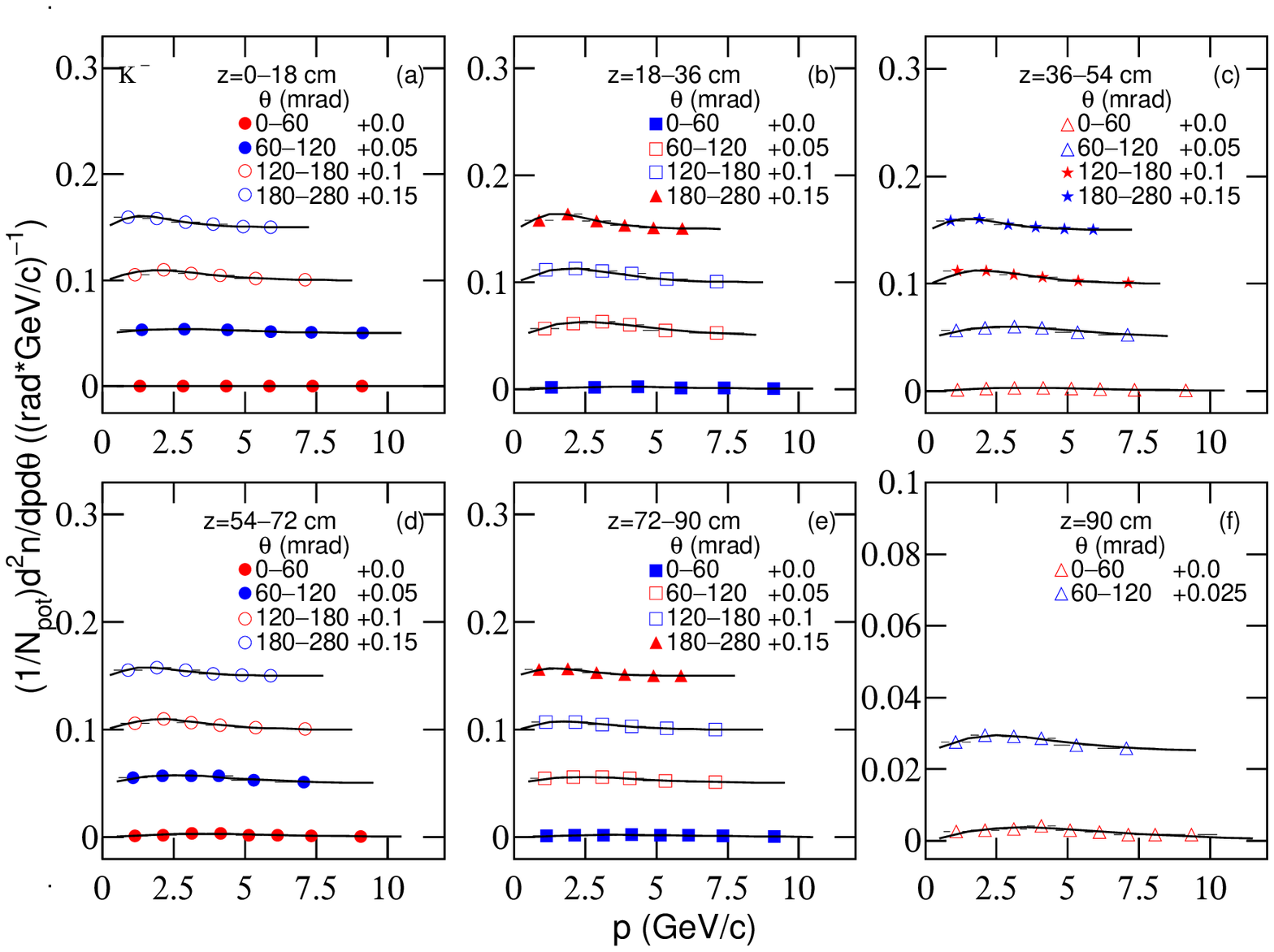}
\end{center}
{\small Fig. 5. Same as Fig. 1, but showing the spectra of $K^-$.
Panels (a), (b), (c), (d), (e), and (f) represent the spectra for
$z=0$--18, 18--36, 36--54, 54--72, 72--90, and 90 cm,
respectively.}
\end{figure*}

\begin{figure*}[!htb]
\begin{center}
\includegraphics[width=12.50cm]{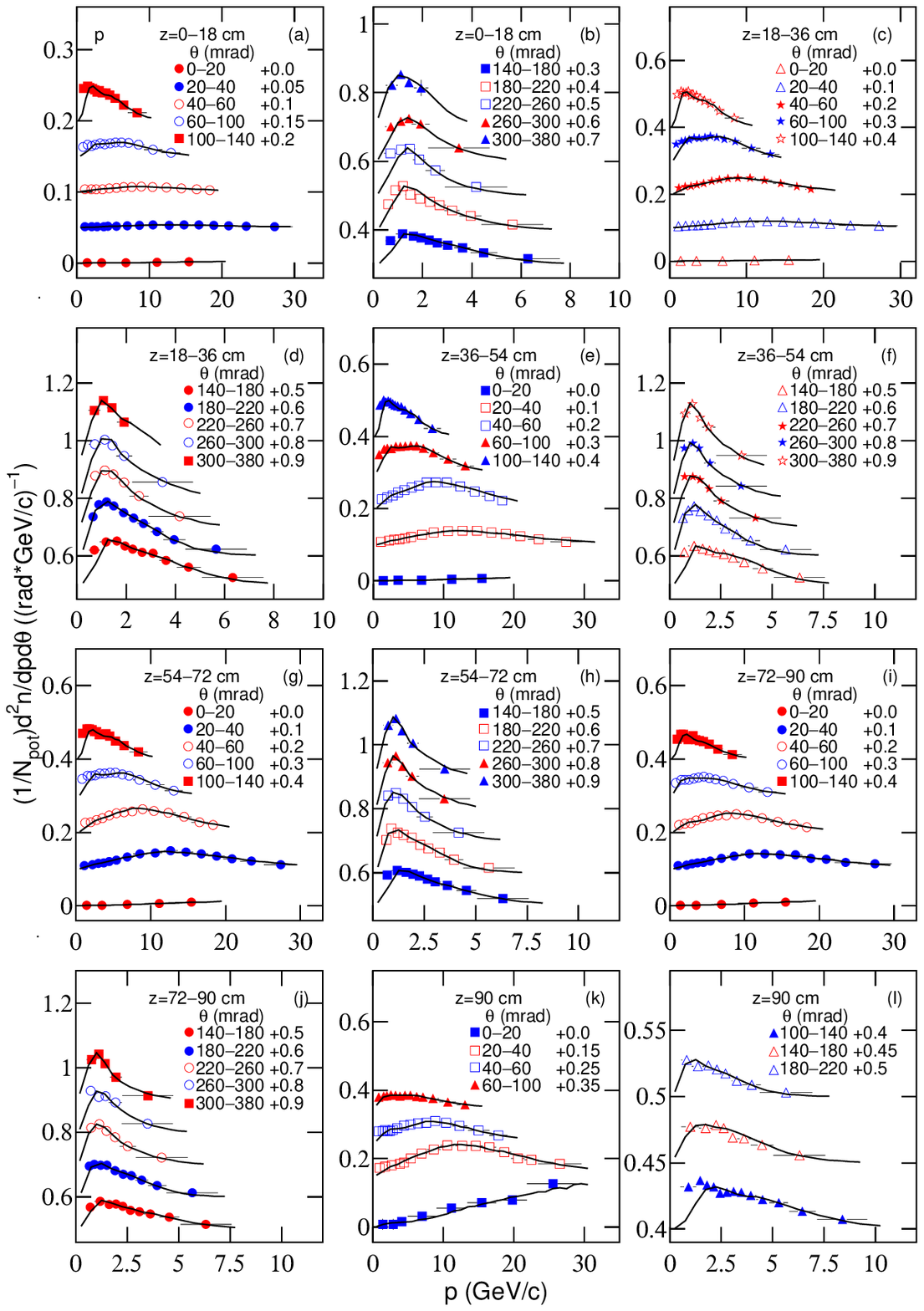}
\end{center}
{\small Fig. 6. Same as Fig. 1, but showing the spectra of $p$.
Panels (a)--(b), (c)--(d), (e)--(f), (g)--(h), (i)--(j), and
(k)--(l) represent the spectra for $z=0$--18, 18--36, 36--54,
54--72, 72--90, and 90 cm, respectively.}
\end{figure*}

\begin{figure*}[!htb]
\begin{center}
\includegraphics[width=12.50cm]{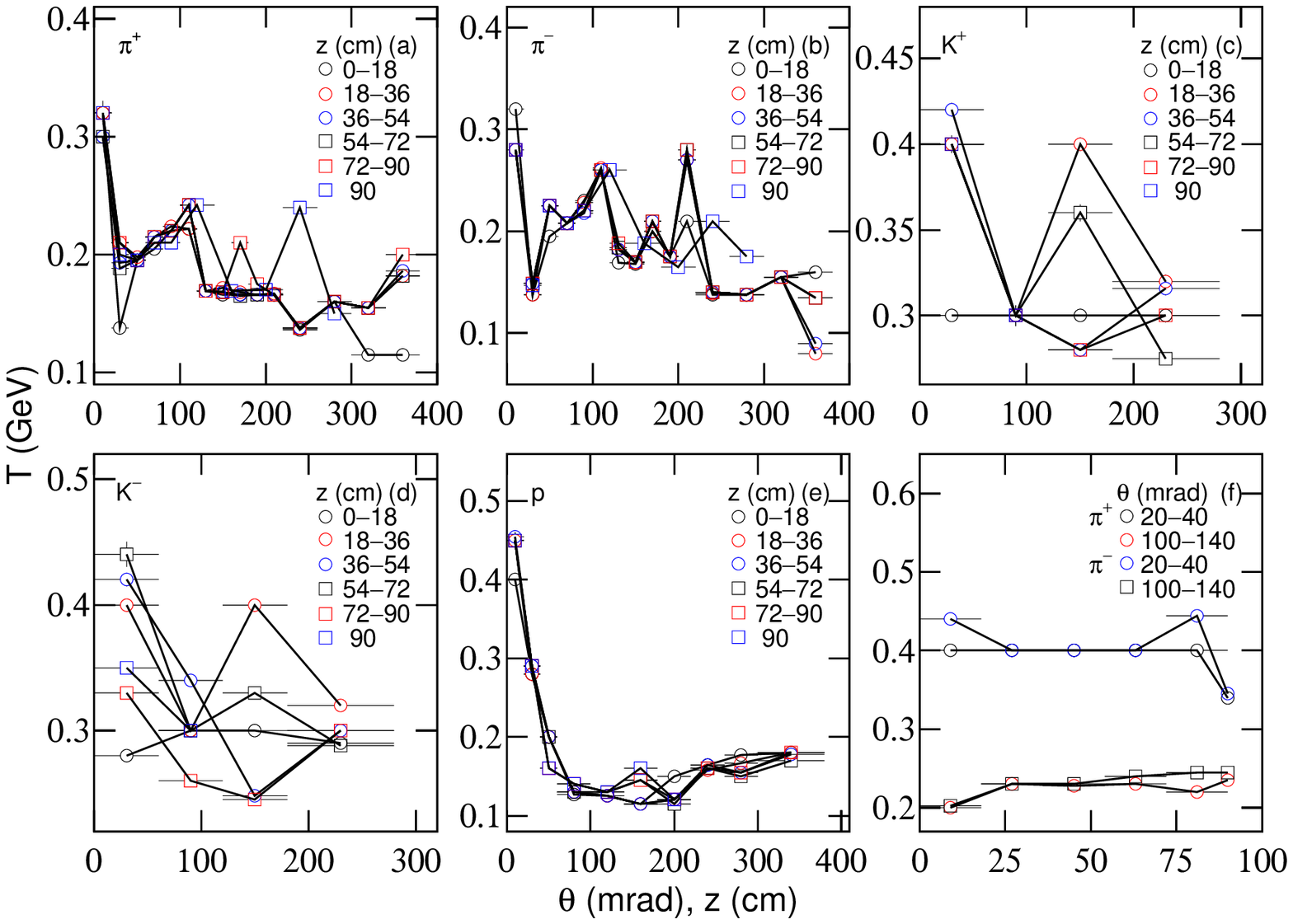}
\end{center}
{\small Fig. 7. Dependence of $T$ on (a)--(e) $\theta$, which are
extracted from the data samples within different $z$ ranges for
$\pi^+$, $\pi^-$, $K^+$, $K^-$, and $p$ respectively, and on (f)
$z$, which are extracted from the data samples within different
$\theta$ ranges for $\pi^+$ and $\pi^-$.}
\end{figure*}

\begin{figure*}[!htb]
\begin{center}
\includegraphics[width=12.50cm]{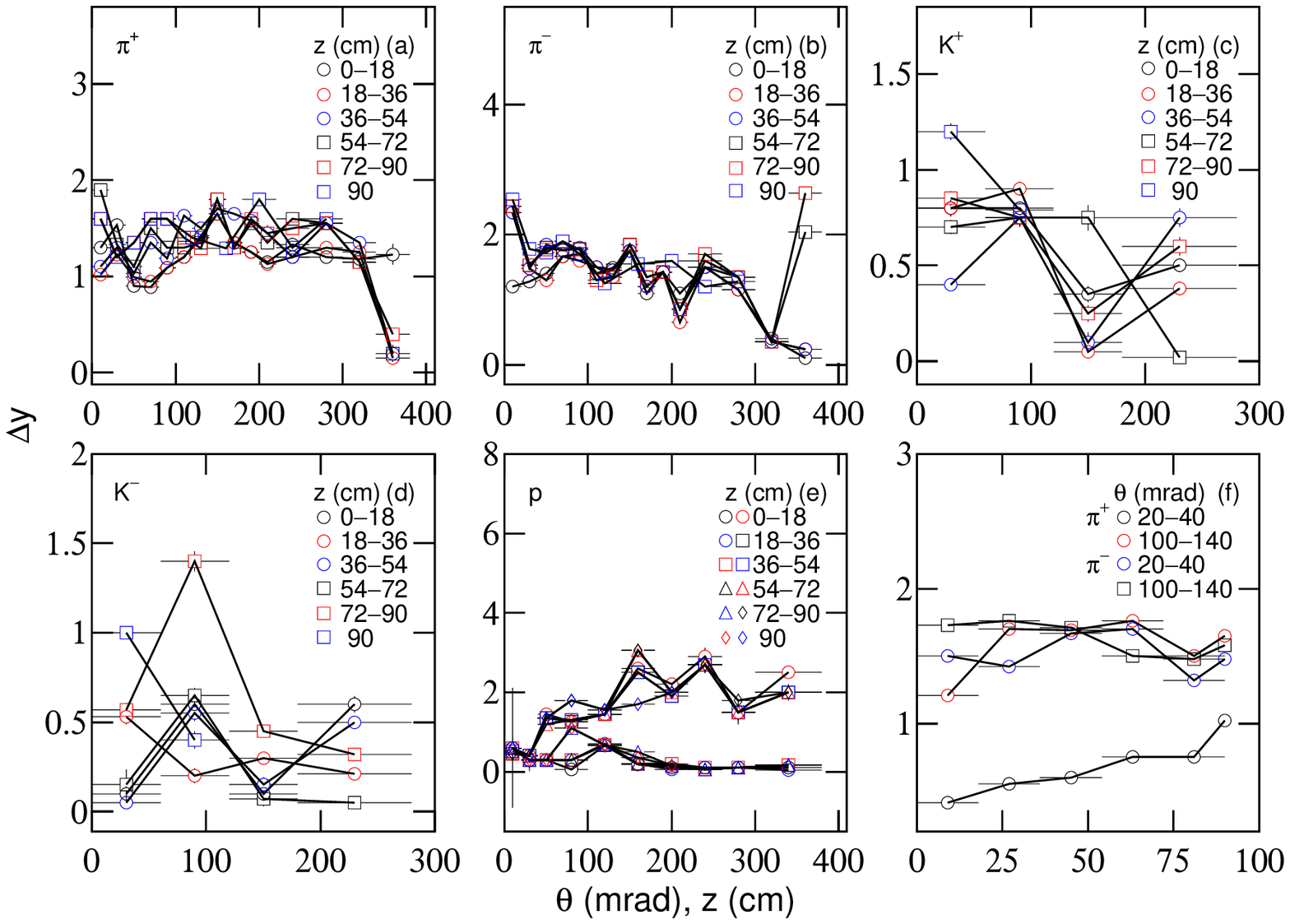}
\end{center}
{\small Fig. 8. Same as Fig. 7, but showing the dependence of
$\Delta y$. Large $\Delta y$ ($=y_{L\max}-y_{L\min}>1$) in panel
(e) represent mainly the rapidity shifts for leading protons.}
\end{figure*}

\begin{figure*}[!htb]
\begin{center}
\includegraphics[width=12.50cm]{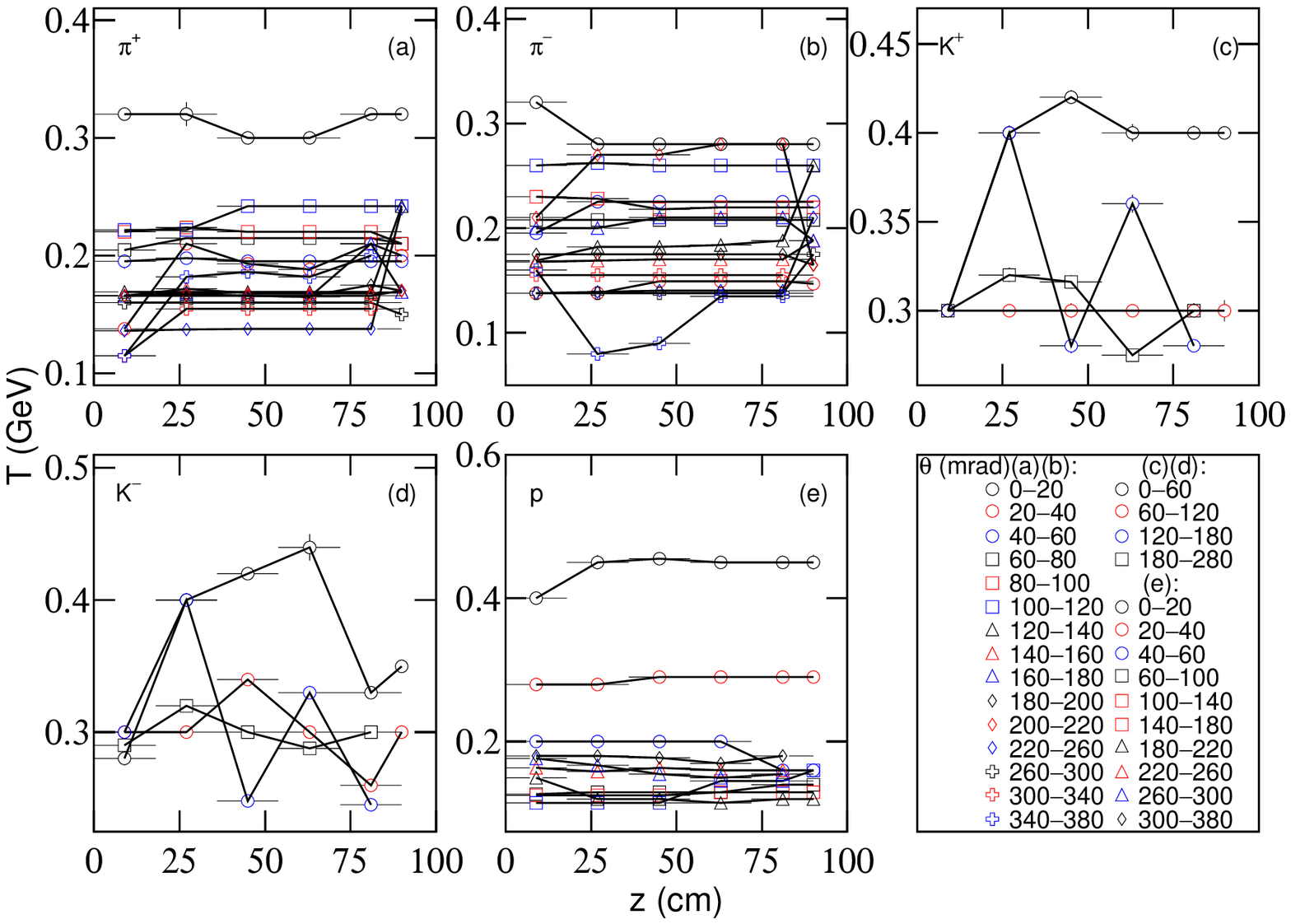}
\end{center}
{\small Fig. 9. Dependence of $T$ on $z$, which are extracted from
the data samples within different $\theta$ ranges for (a) $\pi^+$,
(b) $\pi^-$, (c) $K^+$, (d) $K^-$, and (e) $p$.}
\end{figure*}

\begin{figure*}[!htb]
\begin{center}
\includegraphics[width=12.50cm]{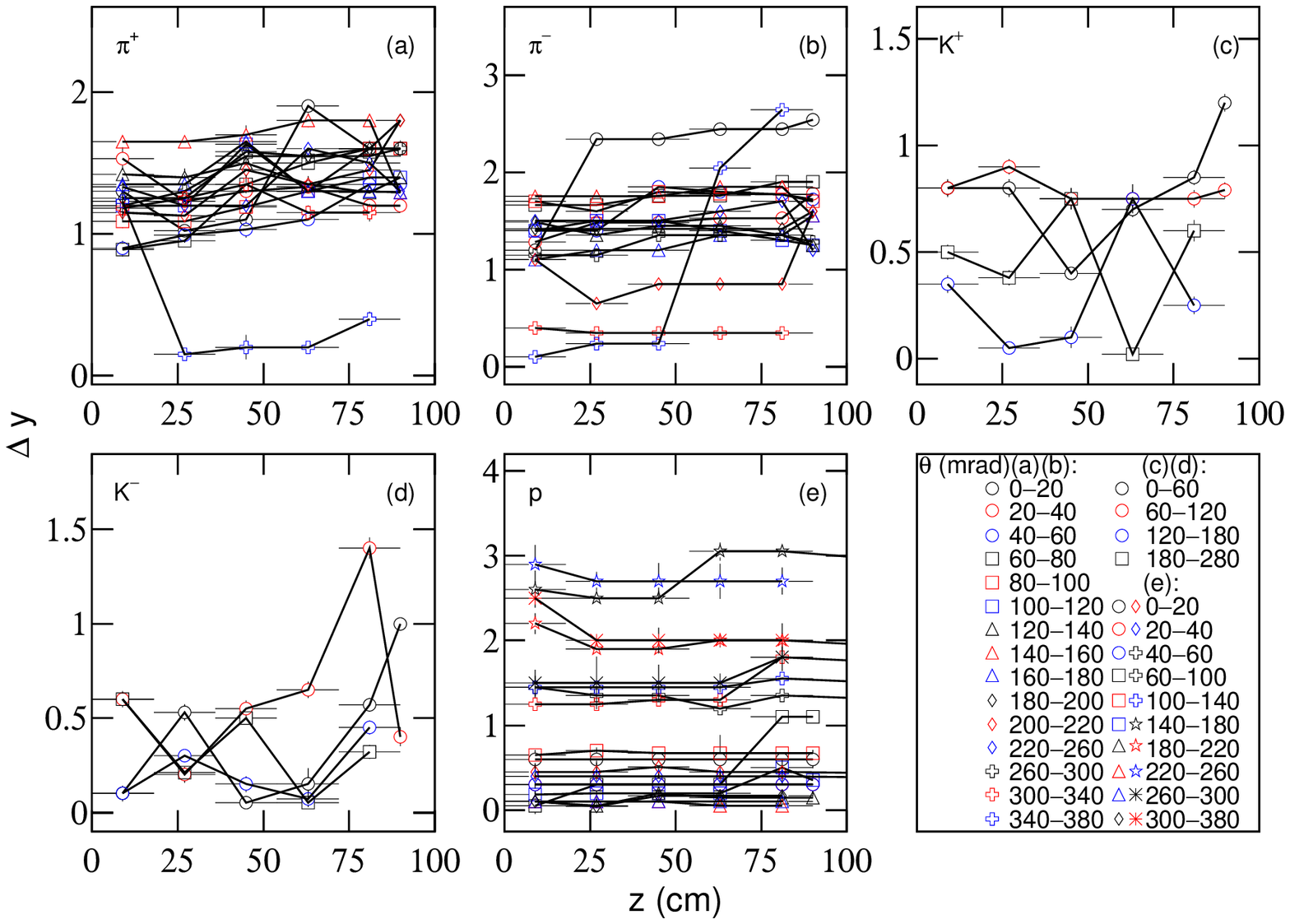}
\end{center}
{\small Fig. 10. Same as Fig. 9, but showing the dependence of
$\Delta y$. Large $\Delta y$ ($=y_{L\max}-y_{L\min}>1$) in panel
(e) represent mainly the rapidity shifts for leading protons.}
\end{figure*}

\begin{figure*}[!htb]
\begin{center}
\includegraphics[width=12.50cm]{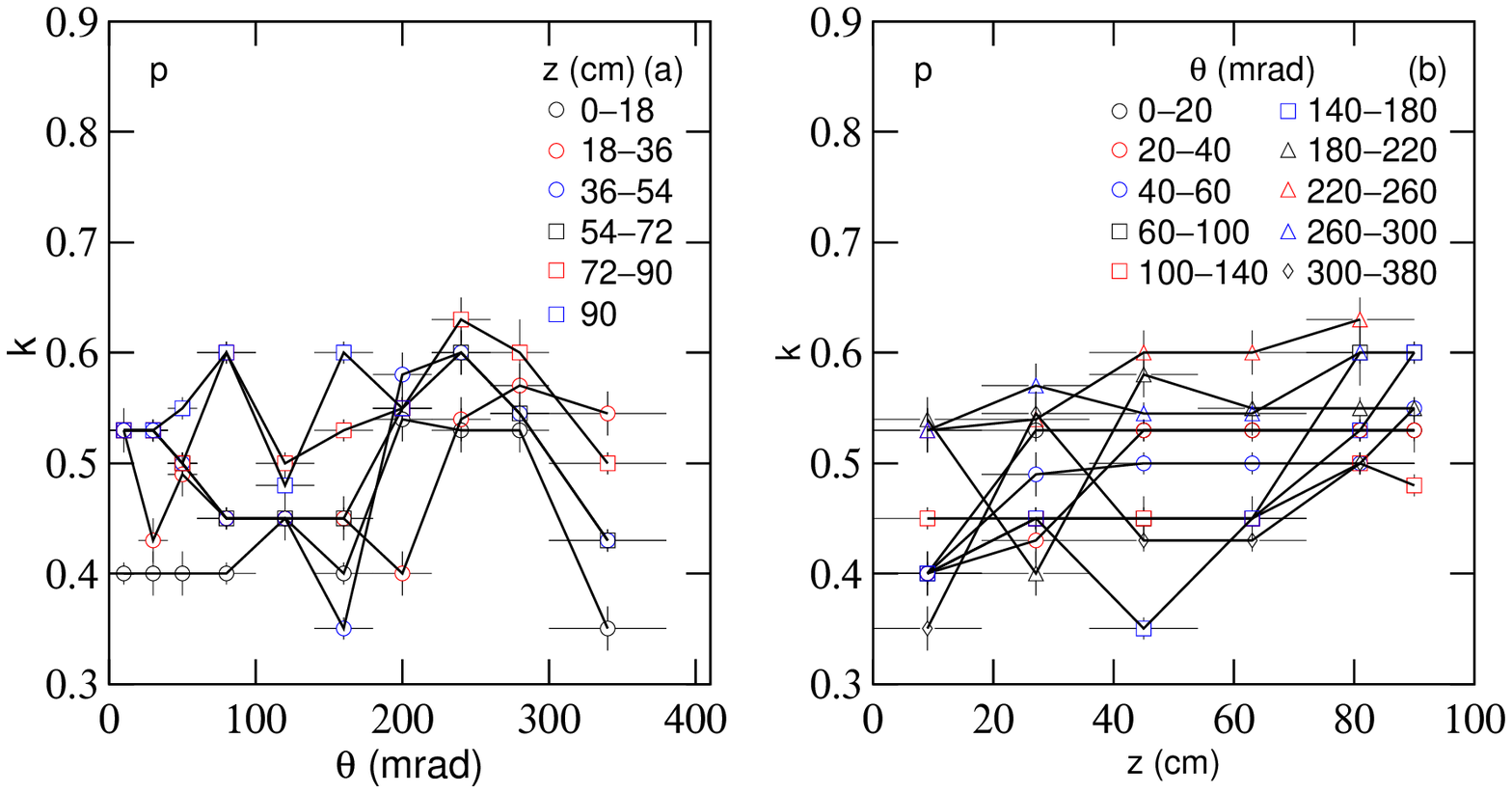}
\end{center}
{\small Fig. 11. Dependences of $k$ on (a) $\theta$ and (b) $z$,
which are extracted from the data samples within different $z$ and
$\theta$ ranges, respectively.}
\end{figure*}

\begin{figure*}[!htb]
\begin{center} \vskip1cm
\includegraphics[width=12.50cm]{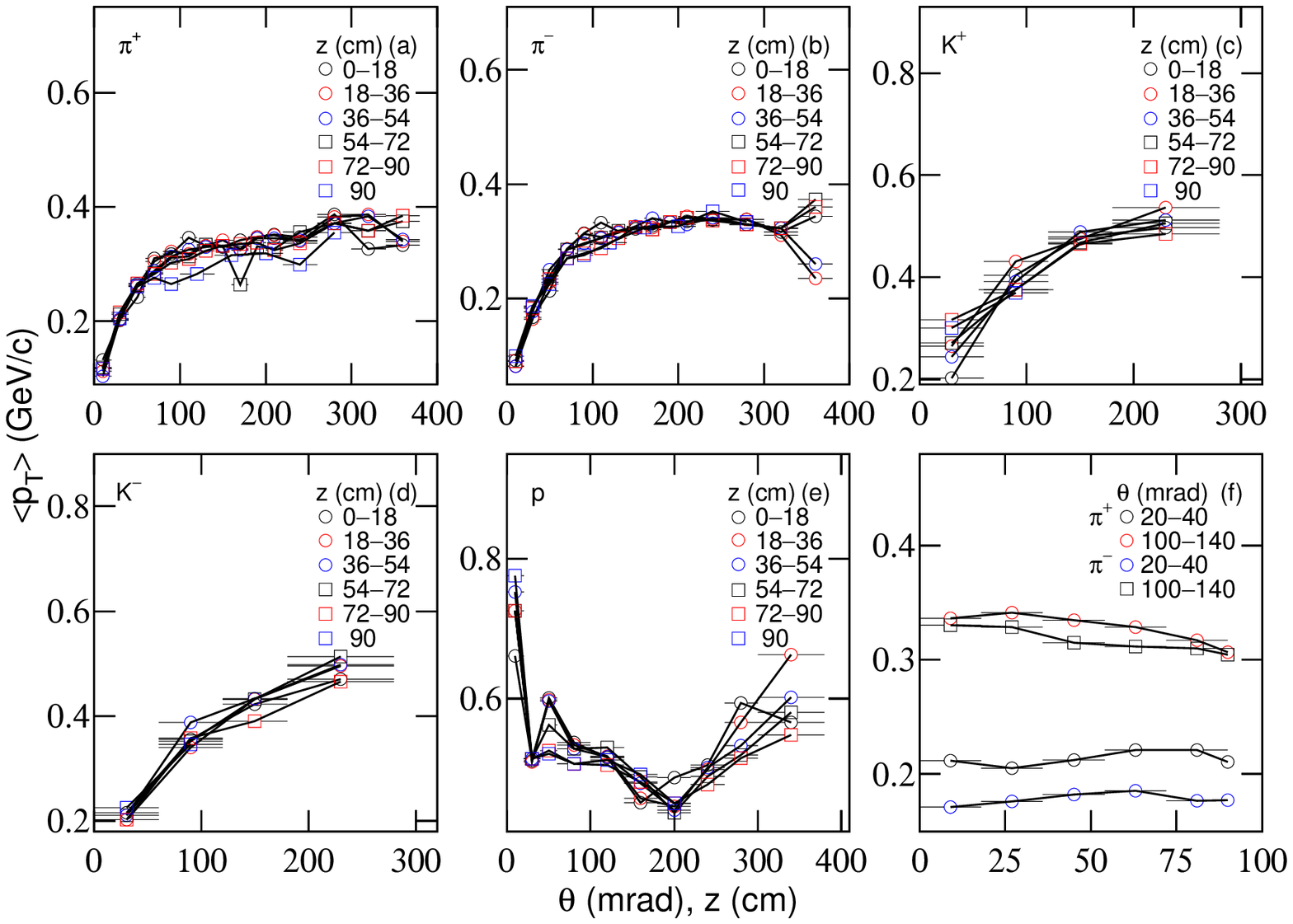}
\end{center}
{\small Fig. 12. Dependence of $\langle p_T\rangle$ on (a)--(e)
$\theta$, which are extracted from the data samples within
different $z$ ranges for $\pi^+$, $\pi^-$, $K^+$, $K^-$, and $p$
respectively, and on (f) $z$, which are extracted from the data
samples within different $\theta$ ranges for $\pi^+$ and $\pi^-$.}
\end{figure*}

\begin{figure*}[!htb]
\begin{center}
\includegraphics[width=12.50cm]{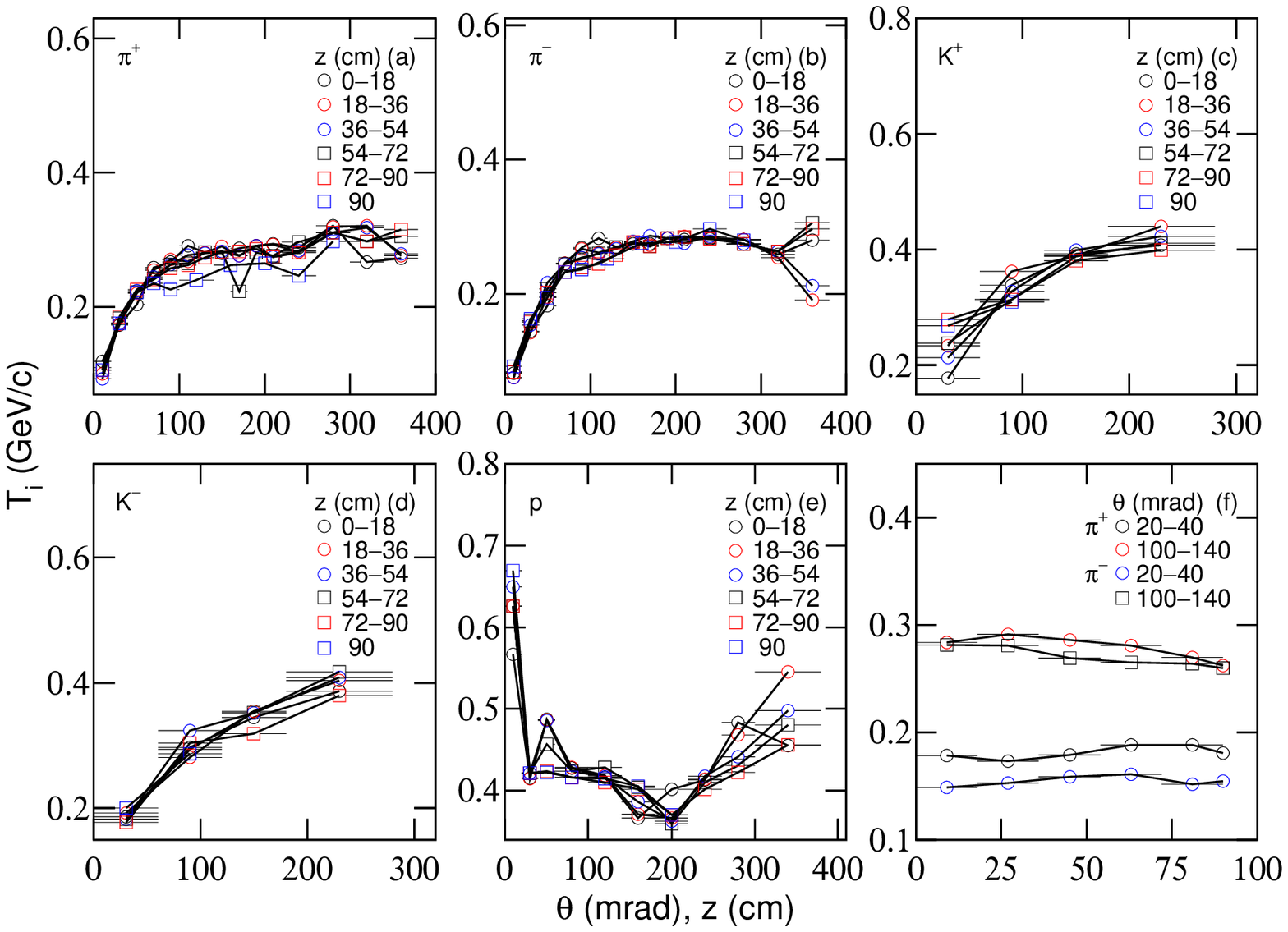}
\end{center}
{\small Fig. 13. Same as Fig. 12, but showing the dependence of
$T_i$.}
\end{figure*}

\begin{figure*}[!htb]
\begin{center} \vskip1cm
\includegraphics[width=12.50cm]{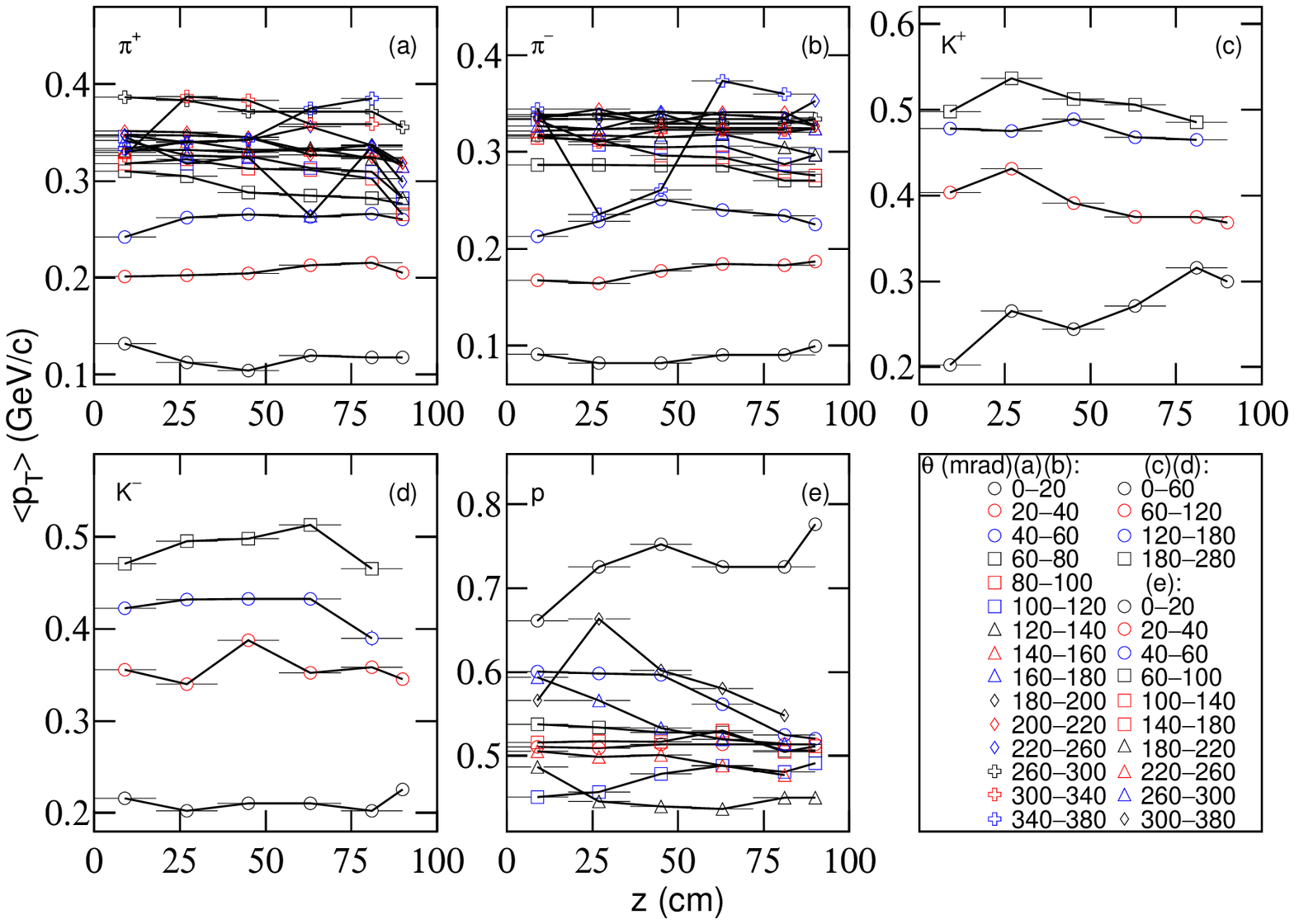}
\end{center}
{\small Fig. 14. Dependence of $\langle p_T\rangle$ on $z$, which
are extracted from the data samples within different $\theta$
ranges for (a) $\pi^+$, (b) $\pi^-$, (c) $K^+$, (d) $K^-$, and (e)
$p$.}
\end{figure*}

\begin{figure*}[!htb]
\begin{center}
\includegraphics[width=12.50cm]{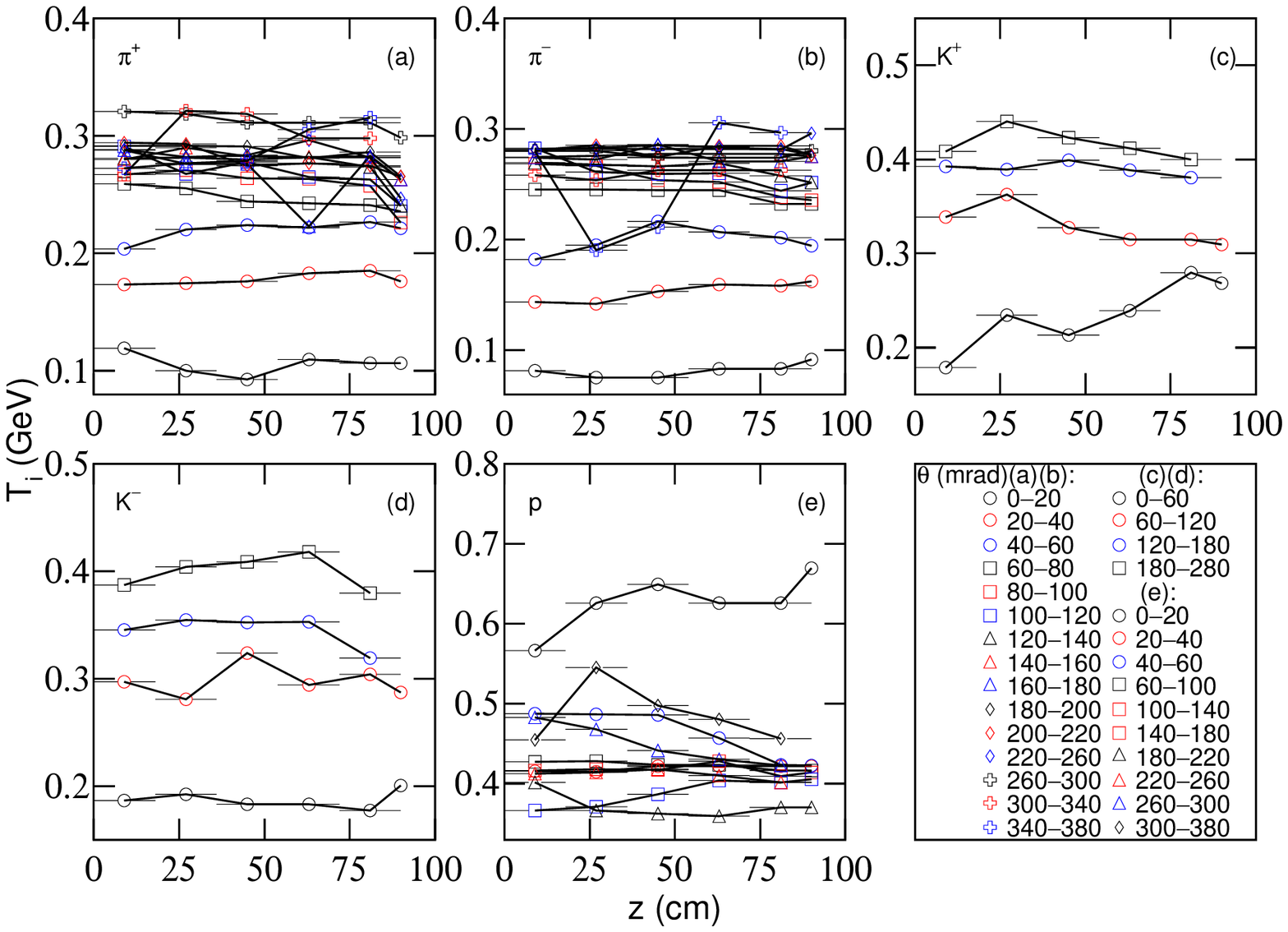}
\end{center}
{\small Fig. 15. Same as Fig. 14, but showing the dependence of
$T_i$.}
\end{figure*}

\begin{figure*}[!htb]
\begin{center}
\includegraphics[width=6.cm]{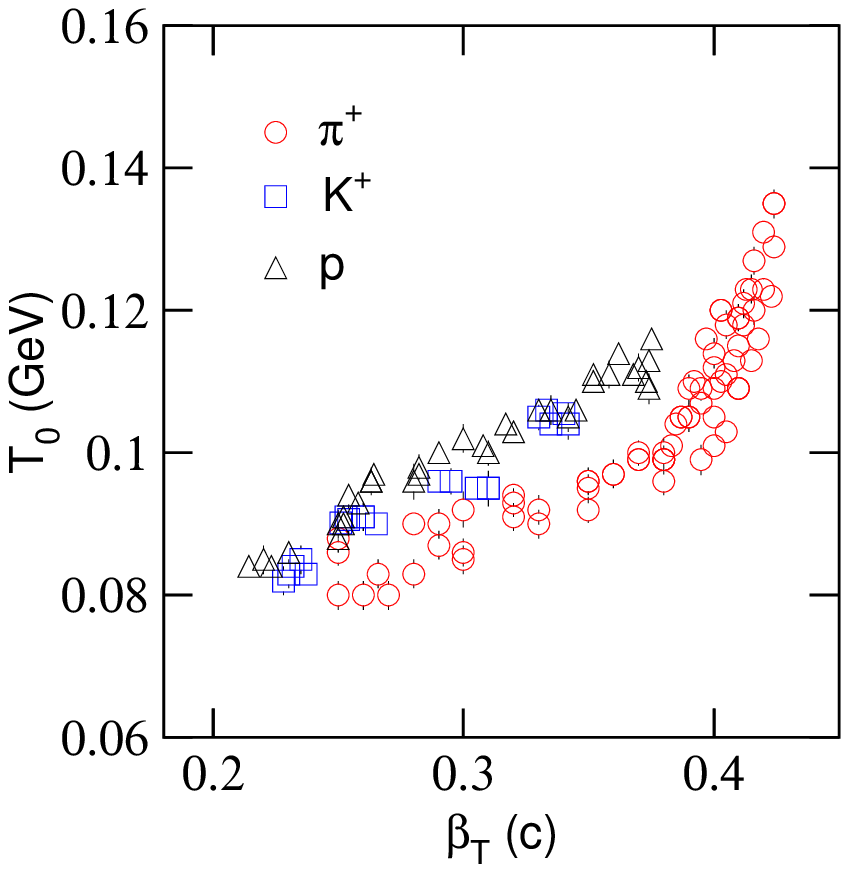}
\end{center}
{\small Fig. 16. Relation between $T_0$ and $\beta_T$ from the
blast-wave model. The symbols represent the results from the
spectra of positive particles.}
\end{figure*}

Similar to Fig. 1, Fig. 6 shows the momentum spectra of $p$
emitted in $p$-C collisions at 31 GeV/$c$. Panels (a)--(b),
(c)--(d), (e)--(f), (g)--(h), (i)--(j), and (k)--(l) represent the
spectra for $z=0$--18, 18--36, 36--54, 54--72, 72--90, and 90 cm,
respectively. The values of $T$, $k$, $y_{\max}$, $y_{\min}$,
$y_{L\max}$, $y_{L\min}$, $N_0$, $\chi^2$, and ndof corresponding
to the fits for the spectra are listed in Table A6 in the
appendix. In a few cases, ndof are negative which appear in the
table in terms of ``$-$" and the corresponding curves are just for
eye guiding only. It should be noted that the contributions of
leading protons have to be considered in the spectra. One can see
that the theoretical model results are approximately in agreement
with the experimental data.

We notice from Tables A1--A6 that different $T$ for a range of $z$
and its dependence with $\theta$ or $y$ are observed, but the
development of the model in our previous work~\cite{20b} concludes
that $T$ is independent of $y$. We would like to explain here that
this paper treats $T$ as differential function of $\theta$ or $y$,
which is more detailed. While, our previous work treats $T$ as
integral or mean quantity over $y$. As for which case should be
used, it depends on experimental data.

We now analyze the dependences of free parameters on $\theta$ and
$z$. Figures 7 and 8 show respectively the dependences of $T$ and
$\Delta y$ ($=y_{\max}-y_{\min}$) on (a)--(e) $\theta$, which are
extracted from the data samples within different $z$ ranges for
$\pi^+$, $\pi^-$, $K^+$, $K^-$, and $p$ respectively, and on (f)
$z$, which are extracted from the data samples within different
$\theta$ ranges for $\pi^+$ and $\pi^-$, where we use $\Delta y$
to denote the difference between $y_{\max}$ and $y_{\min}$ to
avoid trivialness in using both $y_{\max}$ and $y_{\min}$. In
particular, in Fig. 8(e), the results with $\Delta y>1$ are mainly
for leading protons and obtained by $y_{L\max}-y_{L\min}$. One can
see that, for $\pi^{\pm}$ and $K^{\pm}$, $T$ and $\Delta y$
decrease slightly with the increase of $\theta$, and do not change
obviously with the increase of $z$. The obtained $T$ ($\Delta y$)
values for negative and positive pions or kaons seem to be very
similar as we expect. The data for antiproton ($\overline{p}$) are
not available in ref.~\cite{17}, which forbids in making a
comparison for $p$ and $\overline{p}$ in this paper. In fact, the
situation for $p$ is more complex due to the effect of leading
protons.

The dependences of $T$ and $\Delta y$ on $\theta$ for the
productions of $\pi^{\pm}$ and $K^{\pm}$ can be explained by the
effect of cascade collisions in the target and by the nuclear
stopping of the target. The cascade collisions can cause larger
$\theta$ and more energy loss and then lower $T$. The nuclear
stopping can cause smaller $\Delta y$. Combining with cascade
collisions and nuclear stopping, one can obtain low $T$ and small
$\Delta y$ at large $\theta$ for the productions of $\pi^{\pm}$
and $K^{\pm}$. Because of the effect of leading particles, the
situation for the emissions of $p$ is more complex, which shows
different trends from those of $\pi^{\pm}$ and $K^{\pm}$.
Meanwhile, the flow effect can cause larger $T$, which is related
to more complex mechanism.

The dependences of $T$ and $\Delta y$ on $z$, which are extracted
from the data samples within different $\theta$ ranges for (a)
$\pi^+$, (b) $\pi^-$, (c) $K^+$, (d) $K^-$, and (e) $p$, are given
in Figs. 9 and 10 respectively. In particular, large $\Delta y$
($=y_{L\max}-y_{L\min}>1$) in Fig. 10(e) represent mainly the
rapidity shifts of leading protons. In principle, there is no
obvious increase or decrease in $T$ and $\Delta y$ with the
increase of $z$, but some statistical fluctuations in few cases.
This result is natural due to the fact that $z$ is not the main
factor in a 90-cm-long graphite target. It is expected that $T$
and $\Delta y$ will decrease with the increase of $z$ in a very
long graphite target in which the energy loss of the beam protons
has to be considered. The NA61/SHINE experimental data analyzed in
this paper are not obtained from a long graphite target and hence
it is not necessary to consider the energy loss of the beam
protons.

Figure 11 displays the dependences of fraction $k$ of non-leading
protons on (a) $\theta$ and (b) $z$, which are extracted from the
data samples within different $z$ and $\theta$ ranges,
respectively. One can see that there is no obvious change in the
dependence of $k$ on $\theta$, but some statistical fluctuations.
There is a slight increase in the dependence of $k$ on $z$ with
the increase of $z$, which can be explained by more energy loss of
the beam protons at larger $z$. This energy loss is small in a not
too large $z$ range, which does not affect obviously other free
parameters such as $T$ and $\Delta y$ due to their less
sensitivity at the energy in the $z$ range considered in this
paper. It is natural that the larger (fewer) fraction $k$ ($1-k$)
of protons appears as non-leading (leading) particles at lower
energy or larger $z$. Indeed, the fraction is mainly determined by
the collision energy, and the leading protons are considerable at
the SPS. In fact, the leading protons are those existed in the
projectile with high momentum and small emission angle, but not
the produced protons. With the increase of collision energy up to
dozens of GeV and above at which meson-dominated final states
appear~\cite{20c}, $k$ will increase due to the increase of
accompanied produced protons. With the decrease of collision
energy down to several GeV and below at which baryon-dominated
final states appear~\cite{20c}, $k$ will also increase due to the
increase of target stopping which causes the decrease of leading
protons.

Figures 12 and 13 show respectively the dependences of average
$p_T$ ($\langle p_T \rangle$) and $T_i$ on (a)--(e) $\theta$,
which are extracted from the data samples within different $z$
ranges for $\pi^+$, $\pi^-$, $K^+$, $K^-$, and $p$ respectively,
and on (f) $z$, which are extracted from the data samples within
different $\theta$ ranges for $\pi^+$ and $\pi^-$, where $T_i$
denotes the initial quasi-temperature which is given by the
root-mean-square $p_T$ ($\sqrt{\langle p_T^2\rangle}$) over
$\sqrt{2}$ ($\sqrt{\langle p_T^2\rangle/2}$) according to the
color string percolation model~\cite{21,22,23}. It should be noted
that $\sqrt{\langle p_T^2\rangle/2}$ in Refs. ~\cite{21,22,23} is
regarded as the initial temperature. In that model there are free
parameters associated to the medium created in a high energy
collision, which is not the case for this paper at low energy. So
we call $\sqrt{\langle p_T^2\rangle/2}$ the initial
quasi-temperature in this paper. The dependences of $\langle p_T
\rangle$ and $T_i$ on $z$ are presented in Figs. 14 and 15
respectively, which are extracted from the data samples within
different $\theta$ ranges. One can see that, for pions and kaons,
there are increases in $\langle p_T \rangle$ and $T_i$ when
$\theta$ increases. The situation is complex for protons due to
the effect of leading protons which have high momenta and result
in high $\langle p_T \rangle$ and $T_i$ at small $\theta$. The
produced protons which are non-leading should have similar trend
in $\langle p_T \rangle$ and $T_i$ as those for pions and kaons.
As a combination, the final protons are the sum of leading and
produced protons. There is no obvious change in $\langle p_T
\rangle$ and $T_i$ when $z$ increases due to not too large energy
loss in a 90-cm-long graphite target.

We would like to point out that there are different
definitions~\cite{23b} for leading particles in experiments. There
are at least four production mechanisms~\cite{23c,23d} for leading
protons in electron induced deep-inelastic scattering on proton.
Among these mechanisms, at HERA energy, diffractive deep-inelastic
scattering~\cite{23e,23f} in which 72\% of leading protons have
momentum being larger than 0.9$p_{Lab}$ occupy about
26\%~\cite{23c} of leading protons, which are not enough to cover
all leading protons. In particular, for leading protons with
momenta being (0.5--0.98)$p_{Lab}$, a large fraction (77\%) comes
from non-diffractive deep-inelastic scatterings. In proton-proton
and proton-nucleus collisions at the considered energy of this
paper, the fraction of diffractive process is about
20\%~\cite{23g} in inelastic events, which is only a half of the
fraction of leading protons. Even in nucleus-nucleus collisions,
the effect of leading protons in forward rapidity region is also
obvious~\cite{20b,23h,23i,23j}, which also reflects in high
momentum region and is not only from diffractive process.

Naturally, there are other additional arguments to explain the
behavior of Figs. 12 and 13 for the proton case. In fact, there
are multiple or cascade secondary scatterings among produced
particles and target nucleons. As low mass particles, the emission
angles of pions and kaons increase obviously after multiple
scatterings. This results in large $\langle p_T \rangle$ and $T_i$
due to large $\theta$ for pions and kaons. Contrary to this, the
emission angles of protons increase in smaller amount after
multiple scatterings due to higher mass of protons compared to
pions and kaons. This results in small $\langle p_T \rangle$ and
$T_i$ due to small $\theta$ for protons. However, non-negligible
leading protons which have high momenta and smaller angles do not
experience much multiple scatterings, which renders large $\langle
p_T \rangle$ and $T_i$ at small $\theta$. As a competitive result,
protons present different case from pions and kaons.

One can see naturally the coincident trend for $\langle p_T
\rangle$ and $T_i$ in different $\theta$ and $z$ ranges. Due to
the flow effect not being excluded, the trend of $T$ is
inconsistent with that of $T_i$. As an all-around result, the
effects of transverse and longitudinal flows are complex. The flow
effect can obviously affect $T$, which is model dependent. The
flow effect also affects $\langle p_T \rangle$ and $T_i$ which are
also model dependent. Therefore, we mention here that $T$ is not a
``real" temperature, but the effective temperature. In our
opinion, the temperature and flow velocity should be independent
of models, which is usually not the case more often, as some
formalisms are used to extract the radial flow and the
real/thermal temperature, which estimate the real temperature of
the system being dependent of models.

The experimental data cannot be clearly distinguished into two
parts: One part is the contribution of thermal motion, which
reflects the ``real" temperature at the kinetic freeze-out. The
other part is the contribution of the collective flow. The current
blast-wave model~\cite{24,25} treats the thermal motion and flow
effect by using the kinetic freeze-out temperature and transverse
flow velocity, respectively. After fitting the spectra with ${\rm
ndof}>1$ and using $p_T$ coverage as widely as possible
($p_T=0$--3 GeV/$c$), our study using the blast-wave model with
flow profile parameter being 2 can obtain similar fit results as
the curves in Figs. 1--6. To protrude the fit results of thermal
model, the fit results of blast-wave model are not displayed in
these figures. The relation between $T_0$ and $\beta_T$ for
different cases from the spectra of positive particles are plotted
in Fig. 16, where the circles, squares, and triangles represent
the results from $\pi^+$, $K^+$, and $p$ spectra, respectively.
One can see considerable flow-like effect in $p$-C collisions at
31 GeV/$c$, which shows a positive correlation between $T_0$ and
$\beta_T$. The kinetic freeze-out temperature $T_0$ is about from
0.080 to 0.135 GeV. The corresponding transverse flow velocity
$\beta_T$ is about from $0.21$ to $0.42c$. Massive particles such
as $p$ correspond to larger $T_0$ and smaller $\beta_T$ comparing
to $\pi^+$ at the same or similar $\theta$, which is in agreement
with hydrodynamic type behavior. The flow-like effect observed in
this work is slightly less than the flow velocity ($0.3c$ in
peripheral and $0.5c$ in central gold-gold collisions) obtained
from the yield ratio of $p/\pi$ in a simple afterburner
model~\cite{23k}. The difference is due to the fact that lower
energy small system with minimum-bias sample is studied in this
paper. In some cases, the results on kinetic freeze-out
temperature or transverse flow velocity obtained from different
models are not always harmonious~\cite{26,27}.

It should be noted that there is entanglement in determining $T_0$
and $\beta_T$. For a give $p_T$ spectrum, $T_0$ and $\beta_T$ are
negatively correlated, which means an increase in $T$ should
result in a decrease of $\beta_T$. But for a set of $p_T$ spectra,
after determining $T_0$ and $\beta_T$ for each $p_T$ spectrum, the
correlation between $T_0$ and $\beta_T$ is possibly positive or
negative, which depends on the choices of flow profile function
and $p_T$ coverage. If the correlation is negative, one may
increase $T_0$ and decrease $\beta_T$ by changing the flow profile
function and $p_T$ coverage, and obtain possibly positive
correlation. If the correlation is positive, one may decrease
$T_0$ and increase $\beta_T$ by changing the flow profile function
and $p_T$ coverage, and obtain possibly negative correlation.
Unlike experimental papers, where one finds a single $T_0$ and a
common $\beta_T$ by fitting the blast-wave model to the bulk part
of the $p_T$ spectra (in a very narrow coverage which is particle
dependent and much less than 3 GeV/$c$) by performing a
simultaneous fitting to the identified particle spectra using a
changeable $n_0$ (from 0 to 4.3)~\cite{28a}, here we have
considered a differential freeze-out scenario and have restricted
uniformly the fitting up to 3 GeV/$c$ for different particles and
have used always $n_0=2$. The value of $T_0$ ($\beta_T$) in
positive correlation is larger (less) than that in negative
correlation. Positive correlation means high excitation and quick
expansion, while negative correlation means longer lifetime (lower
excitation) and quicker expansion. In our opinion, although both
positive and negative correlations are available, one needs other
method to check which one is suitable. In fact, positive
correlation in Fig. 16 is in agreement with the alternative method
used in our previous works~\cite{26,27}.

We would rather like to use $\langle p_T \rangle$ directly in the
determination of kinetic freeze-out temperature and transverse
flow velocity. For example, the contribution of one participant in
each binary collision in the Erlang distribution is $\langle p_T
\rangle/2$ which is regarded as effective temperature~\cite{28}
contributed by the thermal motion and flow effect. We could assume
the contribution fraction of the thermal motion to be $k_0$. Then,
the kinetic freeze-out temperature is $k_0\langle p_T \rangle/2$,
and the transverse flow velocity is $(1-k_0)\langle p_T
\rangle/2m_0\overline{\gamma}$, where $\overline{\gamma}$ is the
mean Lorentz factor of the considered particles in the rest frame
of emission source. If we take $k_0 \approx 0.3$ and at large
$\theta$, the obtained kinetic freeze-out temperature (0.05 GeV
for pion emission and 0.10 GeV for proton emission) are in
agreement with those from the blast-wave model~\cite{24,25}, and
transverse flow velocity ($0.2c$ for pion emission and $0.1c$ for
proton emission) are qualitatively in agreement with those from
the blast-wave model~\cite{24,25} and the afterburner
model~\cite{23e}. The treatment of $\langle p_T \rangle/2$ is also
model dependent and in agreement with hydrodynamic type behavior.
In addition, larger $\langle p_T \rangle/2$ results in larger
$T_0$ and $\beta_T$, which shows positive correlation between
$T_0$ and $\beta_T$. The positive correlation in Fig. 16 is also
in agreement with the treatment of $\langle p_T \rangle/2$.

Before the summary and conclusions, we would like to point out
that the kinetic freeze-out temperature and transverse flow
velocity obtained in this paper are mass dependent, which renders
a scenario for multiple kinetic freeze-out (differential
freeze-out) \cite{Thakur:2016boy}. The afterburner
model~\cite{23k} uses a mass independent flow velocity, which
renders a scenario for single kinetic freeze-out. There are
arguments on the kinetic freeze-out scenario, which is beyond the
focus of this paper, so we shall not discuss it anymore. In
addition, it should be noted that in the absence of required
number of experimental data points, the fittings using the current
model in few cases yield negative $\chi^2$/ndof, making the
description unphysical, though the corresponding curves could be
used as eye guiding only.
\\

{\section{Summary and conclusions}}

We summarize here our main observations and conclusions.

(a) The momentum spectra of $\pi^+$, $\pi^-$, $K^+$, $K^-$, and
$p$ produced in $p$-C collisions at 31 GeV/$c$ are analyzed in the
framework of multisource thermal model by using the Boltzmann
distribution and Monte Carlo method. The results are approximately
in agreement with the experimental data in various emission angle,
$\theta$, ranges and longitudinal positions, $z$, measured by the
NA61/SHINE Collaboration at the SPS.

(b) The effective temperature $T$ and rapidity shifts $\Delta y$
from the spectra under given experimental conditions which limit
various $\theta$ and $z$ ranges are obtained. For $\pi^{\pm}$ and
$K^{\pm}$, $T$ and $\Delta y$ decrease slightly with the increase
of $\theta$, and do not change obviously with the increase of $z$.
The situation for $p$ is more complex due to the effect of leading
protons. There is no obvious change in $T$ and $\Delta y$ when $z$
increases due to not too large energy loss in a not too long
graphite target. Both $T$ and $\Delta y$ depend on models. In
particular, $T$ contains the contribution of flow effect, which is
not ideal to describe the excitation degree of emission source.

(c) The fraction $k$ ($1-k$) of non-leading (leading) protons in
total protons from the spectra in various $\theta$ and $z$ ranges
are obtained. There is no obvious change in the dependence of $k$
($1-k$) on $\theta$, but some statistical fluctuations. There is a
slight increase (decrease) in the dependence of $k$ ($1-k$) on $z$
with the increase of $z$ due to more energy loss of the beam
protons in the target at larger $z$. The effect of leading protons
cannot be neglected at the SPS energies. It is expected that $k$
($1-k$) will be larger (smaller) at both lower ($\leq$ several
GeV) and higher energies ($\geq$ dozens of GeV).

(d) The average transverse momentum $\langle p_T \rangle$ and
initial quasi-temperature $T_i$ from the spectra in various
$\theta$ and $z$ ranges are obtained. For $\pi^{\pm}$ and
$K^{\pm}$, there are increases in $\langle p_T \rangle$ and $T_i$
when $\theta$ increases. The situation for $p$ is complex due to
the effect of leading protons. There is no obvious change in
$\langle p_T \rangle$ and $T_i$ when $z$ increases due to not too
large energy loss in a not too long graphite target. Both $\langle
p_T \rangle$ and $T_i$ are model dependent due to the fact that
they are obtained from the model which fits the data.

(e) The behaviors of effective temperature, rapidity shifts,
fraction of non-leading (leading) protons, average transverse
momentum, and initial quasi-temperature obtained from the fits of
multisource thermal model to the NA61/SHINE data can be explained
in terms of cascade collisions in the target, stopping power of
the target, energy loss of the beam protons in the target, and so
on. This paper provides a new evidence for the effectiveness of
the multisource thermal model, though there is no connection with
a possible formation of a Quark-Gluon Plasma due to small system
being considered.
\\
\\
{\bf Data Availability}

The data used to support the findings of this study are included
within the article and are cited at relevant places within the
text as references.
\\
\\
{\bf Compliance with Ethical Standards}

The authors declare that they are in compliance with ethical
standards regarding the content of this paper.
\\
\\
{\bf Conflict of Interest}

The authors declare that there are no conflicts of interest
regarding the publication of this paper.
\\
\\
{\bf Acknowledgments}

Communications from Debasish Das are highly acknowledged. Authors
P.P.Y., M.Y.D., and F.H.L. acknowledge the financial supports from
the National Natural Science Foundation of China under Grant Nos.
11575103 and 11847311, the Shanxi Provincial Innovative Foundation
for Graduate Education Grant No. 2019SY053, the Scientific and
Technological Innovation Programs of Higher Education Institutions
in Shanxi (STIP) under Grant No. 201802017, the Shanxi Provincial
Natural Science Foundation under Grant No. 201701D121005, and the
Fund for Shanxi ``1331 Project" Key Subjects Construction. Author
R.S. acknowledges the financial supports from ALICE Project No.
SR/MF/PS-01/2014-IITI(G) of Department of Science \& Technology,
Government of India. The funding agencies have no role in the
design of the study; in the collection, analysis, or
interpretation of the data; in the writing of the manuscript, or
in the decision to publish the results.
\\

\begin{table*}[!htb]
{\bf Appendix: The tables for parameters}
\\
\\

{\scriptsize Table A1. Values of $T$, $y_{\max}$, $y_{\min}$,
$N_0$, $\chi^2$, and ndof corresponding to the curves in Fig. 1 in
which different data are measured in different $\theta$ and $z$
ranges. In the table, $z$ is in the units of cm, and $\theta$ is
not listed, which appears in Fig. 1. In one case, ndof is negative
which appears in terms of ``$-$" and the corresponding curve is
just for eye guiding purpose.} {\tiny
\begin{center} \vskip-0.3cm
\begin{tabular} {cccccccccccc}\\ \hline\hline Figure &
  $T$ (GeV) & $y_{\max}$& $y_{\min}$& $N_0(\times0.001)$ & $\chi^2$/ndof \\
\hline
                 & $0.320\pm0.005$ & $2.30\pm0.02 $ & $1.10\pm0.02 $ & $0.143  \pm0.010 $ & 22/2 \\
  Fig. 1(a)      & $0.138\pm0.004$ & $3.25\pm0.05 $ & $1.72\pm0.06 $ & $0.962  \pm0.020 $ & 42/12\\
0$\leq$$z$$<$18  & $0.195\pm0.006$ & $2.47\pm0.04 $ & $1.57\pm0.03 $ & $2.918  \pm0.100 $ & 93/12\\
                 & $0.205\pm0.003$ & $2.36\pm0.04 $ & $1.47\pm0.03 $ & $6.431  \pm0.200 $ & 87/9 \\
                 & $0.220\pm0.005$ & $2.09\pm0.04 $ & $1.00\pm0.03 $ & $11.108 \pm0.400 $ & 85/9 \\
 \hline
                 & $0.222\pm0.004$ & $2.00\pm0.03 $ & $0.80\pm0.04 $ & $14.541 \pm0.300 $ & 60/9 \\
  Fig. 1(b)      & $0.169\pm0.003$ & $2.20\pm0.04 $ & $0.78\pm0.02 $ & $16.701 \pm0.340 $ & 15/9 \\
0$\leq$$z$$<$18  & $0.166\pm0.003$ & $2.10\pm0.03 $ & $0.45\pm0.02 $ & $17.572 \pm0.260 $ & 16/6 \\
                 & $0.166\pm0.002$ & $2.00\pm0.03 $ & $0.65\pm0.03 $ & $19.149 \pm0.300 $ & 37/6 \\
                 & $0.166\pm0.002$ & $1.90\pm0.02 $ & $0.65\pm0.02 $ & $18.495 \pm0.340 $ & 57/6 \\
 \hline
                 & $0.166\pm0.001$ & $1.80\pm0.02 $ & $0.65\pm0.02 $ & $17.839 \pm0.200 $ & 70/6 \\
  Fig. 1(c)      & $0.136\pm0.002$ & $2.08\pm0.06 $ & $0.75\pm0.04 $ & $31.163 \pm0.720 $ & 35/3 \\
0$\leq$$z$$<$18  & $0.160\pm0.003$ & $1.75\pm0.03 $ & $0.55\pm0.02 $ & $25.050 \pm0.560 $ & 28/2 \\
                 & $0.115\pm0.004$ & $2.08\pm0.08 $ & $0.90\pm0.05 $ & $26.483 \pm0.800 $ & 12/1 \\
                 & $0.115\pm0.004$ & $2.08\pm0.04 $ & $0.85\pm0.10 $ & $25.278 \pm1.200 $ & 2/$-$ \\
\hline
                 & $0.320\pm0.010$ & $2.50\pm0.02 $ & $1.48\pm0.01 $ & $0.571  \pm0.020 $ & 49/3 \\
  Fig. 1(d)      & $0.210\pm0.004$ & $2.73\pm0.02 $ & $1.50\pm0.04 $ & $5.224  \pm0.080 $ & 48/12\\
18$\leq$$z$$<$36 & $0.198\pm0.003$ & $2.56\pm0.04 $ & $1.57\pm0.03 $ & $17.336 \pm0.400 $ & 76/12\\
                 & $0.215\pm0.004$ & $2.30\pm0.03 $ & $1.35\pm0.02 $ & $28.156 \pm1.000 $ & 90/9 \\
                 & $0.224\pm0.004$ & $2.09\pm0.03 $ & $1.00\pm0.02 $ & $31.294 \pm0.400 $ & 67/9 \\
 \hline
                 & $0.222\pm0.005$ & $1.90\pm0.04 $ & $0.70\pm0.03 $ & $31.345 \pm0.800 $ & 70/9 \\
  Fig. 1(e)      & $0.169\pm0.004$ & $2.20\pm0.03 $ & $0.80\pm0.05 $ & $30.855 \pm0.400 $ & 29/9 \\
18$\leq$$z$$<$36 & $0.172\pm0.005$ & $2.10\pm0.02 $ & $0.45\pm0.02 $ & $28.588 \pm0.600 $ & 3/6  \\
                 & $0.168\pm0.002$ & $1.90\pm0.04 $ & $0.55\pm0.05 $ & $25.800 \pm0.400 $ & 19/6 \\
                 & $0.166\pm0.001$ & $1.90\pm0.02 $ & $0.65\pm0.01 $ & $24.237 \pm0.200 $ & 50/6 \\
 \hline
                 & $0.167\pm0.002$ & $1.78\pm0.03 $ & $0.65\pm0.02 $ & $22.902 \pm0.400 $ & 64/6 \\
  Fig. 1(f)      & $0.137\pm0.001$ & $1.95\pm0.02 $ & $0.75\pm0.02 $ & $38.121 \pm0.800 $ & 44/3 \\
18$\leq$$z$$<$36 & $0.160\pm0.002$ & $1.75\pm0.03 $ & $0.45\pm0.03 $ & $33.929 \pm0.800 $ & 22/2 \\
                 & $0.155\pm0.002$ & $1.70\pm0.03 $ & $0.45\pm0.02 $ & $30.541 \pm0.800 $ & 6/1 \\
                 & $0.182\pm0.003$ & $0.70\pm0.02 $ & $0.55\pm0.03 $ & $31.012 \pm0.600 $ & 7/0 \\
\hline
                 & $0.300\pm0.002$ & $2.50\pm0.02 $ & $1.40\pm0.02 $ & $1.064  \pm0.030 $ & 38/3 \\
  Fig. 1(g)      & $0.193\pm0.003$ & $2.85\pm0.01 $ & $1.55\pm0.02 $ & $10.836 \pm0.200 $ & 108/12\\
36$\leq$$z$$<$54 & $0.195\pm0.003$ & $2.60\pm0.05 $ & $1.57\pm0.03 $ & $23.318 \pm0.400 $ & 85/12\\
                 & $0.215\pm0.005$ & $2.30\pm0.03 $ & $0.95\pm0.03 $ & $21.566 \pm0.600 $ & 60/9 \\
                 & $0.220\pm0.003$ & $2.09\pm0.02 $ & $0.90\pm0.02 $ & $27.103 \pm0.400 $ & 69/9 \\
 \hline
                 & $0.242\pm0.002$ & $1.88\pm0.02 $ & $0.25\pm0.03 $ & $26.968 \pm0.600 $ & 42/9 \\
  Fig. 1(h)      & $0.169\pm0.003$ & $2.20\pm0.02 $ & $0.70\pm0.03 $ & $26.177 \pm0.400 $ & 29/9 \\
36$\leq$$z$$<$54 & $0.168\pm0.003$ & $2.10\pm0.04 $ & $0.40\pm0.05 $ & $22.962 \pm0.400 $ & 13/6 \\
                 & $0.166\pm0.005$ & $1.95\pm0.03 $ & $0.30\pm0.02 $ & $21.297 \pm0.600 $ & 40/6 \\
                 & $0.166\pm0.002$ & $1.93\pm0.03 $ & $0.35\pm0.04 $ & $19.830 \pm0.500 $ & 44/6 \\
 \hline
                 & $0.166\pm0.002$ & $1.70\pm0.07 $ & $0.25\pm0.05 $ & $17.778 \pm0.300 $ & 97/6 \\
  Fig. 1(i)      & $0.138\pm0.002$ & $1.95\pm0.03 $ & $0.75\pm0.04 $ & $32.081 \pm0.800 $ & 37/3 \\
36$\leq$$z$$<$54 & $0.160\pm0.003$ & $1.75\pm0.05 $ & $0.20\pm0.04 $ & $23.055 \pm0.600 $ & 5/2 \\
                 & $0.155\pm0.005$ & $1.70\pm0.05 $ & $0.35\pm0.04 $ & $26.196 \pm0.800 $ & 6/1 \\
                 & $0.186\pm0.005$ & $0.70\pm0.07 $ & $0.50\pm0.05 $ & $21.033 \pm0.600 $ & 22/1 \\
\hline
                 & $0.300\pm0.004$ & $2.70\pm0.02 $ & $0.80\pm0.04 $ & $1.561  \pm0.040 $ & 66/3 \\
  Fig. 1(j)      & $0.188\pm0.001$ & $2.93\pm0.02 $ & $1.60\pm0.02 $ & $11.850 \pm0.200 $ & 145/12\\
54$\leq$$z$$<$72 & $0.195\pm0.003$ & $2.60\pm0.03 $ & $1.50\pm0.02 $ & $18.466 \pm0.400 $ & 81/12\\
                 & $0.215\pm0.005$ & $2.30\pm0.03 $ & $0.80\pm0.03 $ & $21.028 \pm0.500 $ & 60/9 \\
                 & $0.220\pm0.004$ & $2.10\pm0.03 $ & $0.80\pm0.02 $ & $21.284 \pm0.400 $ & 72/9 \\
 \hline
                 & $0.242\pm0.005$ & $1.80\pm0.02 $ & $0.50\pm0.03 $ & $19.705 \pm0.300 $ & 76/9 \\
  Fig. 1(k)      & $0.169\pm0.002$ & $2.20\pm0.02 $ & $0.85\pm0.03 $ & $19.159 \pm0.400 $ & 36/9 \\
54$\leq$$z$$<$72 & $0.168\pm0.004$ & $2.10\pm0.04 $ & $0.30\pm0.03 $ & $16.753 \pm0.400 $ & 13/6 \\
                 & $0.165\pm0.001$ & $1.60\pm0.01 $ & $0.30\pm0.01 $ & $14.510 \pm0.200 $ & 87/6 \\
                 & $0.166\pm0.001$ & $1.85\pm0.03 $ & $0.30\pm0.02 $ & $15.317 \pm0.300 $ & 35/6 \\
 \hline
                 & $0.166\pm0.001$ & $1.70\pm0.05 $ & $0.35\pm0.03 $ & $14.737 \pm0.200 $ & 94/6 \\
  Fig. 1(l)      & $0.138\pm0.002$ & $2.20\pm0.02 $ & $0.60\pm0.02 $ & $26.931 \pm0.400 $ & 22/3 \\
54$\leq$$z$$<$72 & $0.160\pm0.002$ & $1.75\pm0.03 $ & $0.20\pm0.03 $ & $23.055 \pm0.400 $ & 5/2 \\
                 & $0.155\pm0.002$ & $1.50\pm0.04 $ & $0.35\pm0.03 $ & $20.234 \pm0.240 $ & 8/1 \\
                 & $0.182\pm0.001$ & $0.85\pm0.02 $ & $0.65\pm0.01 $ & $17.919 \pm0.280 $ & 16/1 \\
\hline
                 & $0.320\pm0.005$ & $2.60\pm0.03 $ & $1.00\pm0.03 $ & $1.791  \pm0.020 $ & 49/3 \\
  Fig. 1(m)      & $0.210\pm0.003$ & $2.80\pm0.02 $ & $1.60\pm0.03 $ & $10.081 \pm0.300 $ & 56/12\\
72$\leq$$z$$<$90 & $0.195\pm0.002$ & $2.65\pm0.03 $ & $1.30\pm0.03 $ & $13.984 \pm0.400 $ & 67/12\\
                 & $0.215\pm0.004$ & $2.30\pm0.02 $ & $0.70\pm0.03 $ & $15.619 \pm0.300 $ & 52/9 \\
                 & $0.220\pm0.003$ & $2.10\pm0.03 $ & $0.50\pm0.03 $ & $15.793 \pm0.400 $ & 66/9 \\
 \hline
                 & $0.242\pm0.003$ & $1.80\pm0.02 $ & $0.40\pm0.02 $ & $15.019 \pm0.300 $ & 64/9 \\
  Fig. 1(n)      & $0.169\pm0.003$ & $2.14\pm0.02 $ & $0.85\pm0.03 $ & $14.177 \pm0.400 $ & 73/9 \\
72$\leq$$z$$<$90 & $0.168\pm0.004$ & $2.10\pm0.03 $ & $0.30\pm0.03 $ & $12.801 \pm0.300 $ & 9/6 \\
                 & $0.210\pm0.003$ & $1.60\pm0.03 $ & $0.30\pm0.02 $ & $11.752 \pm0.200 $ & 14/6 \\
                 & $0.175\pm0.003$ & $1.80\pm0.05 $ & $0.20\pm0.03 $ & $11.763 \pm0.200 $ & 36/6 \\
 \hline
                 & $0.166\pm0.005$ & $1.70\pm0.04 $ & $0.25\pm0.05 $ & $11.174 \pm0.200 $ & 66/6 \\
  Fig. 1(o)      & $0.138\pm0.002$ & $2.00\pm0.03 $ & $0.50\pm0.03 $ & $18.580 \pm0.400 $ & 35/3 \\
72$\leq$$z$$<$90 & $0.160\pm0.002$ & $1.75\pm0.02 $ & $0.20\pm0.02 $ & $18.836 \pm0.200 $ & 13/2 \\
                 & $0.155\pm0.002$ & $1.50\pm0.06 $ & $0.35\pm0.02 $ & $14.886 \pm0.400 $ & 3/1 \\
                 & $0.200\pm0.003$ & $0.85\pm0.04 $ & $0.45\pm0.03 $ & $14.089 \pm0.400 $ & 19/1 \\
\hline
                 & $0.320\pm0.001$ & $2.60\pm0.01 $ & $1.00\pm0.01 $ & $16.231 \pm0.200 $ & 46/3 \\
  Fig. 1(p)      & $0.200\pm0.003$ & $2.80\pm0.02 $ & $1.60\pm0.03 $ & $20.333 \pm0.200 $ & 97/12\\
 $z$$=$90        & $0.195\pm0.005$ & $2.62\pm0.03 $ & $1.30\pm0.03 $ & $16.733 \pm0.200 $ & 88/12\\
                 & $0.210\pm0.004$ & $2.30\pm0.02 $ & $0.70\pm0.03 $ & $13.641 \pm0.300 $ & 50/9 \\
                 & $0.210\pm0.002$ & $2.00\pm0.01 $ & $0.40\pm0.02 $ & $10.621 \pm0.200 $ & 104/9 \\
 \hline
                 & $0.242\pm0.003$ & $1.60\pm0.02 $ & $0.10\pm0.02 $ & $14.572 \pm0.400 $ & 83/9 \\
  Fig. 1(q)      & $0.169\pm0.003$ & $1.80\pm0.03 $ & $0.90\pm0.02 $ & $8.858  \pm0.400 $ & 64/6 \\
 $z$$=$90        & $0.170\pm0.002$ & $1.60\pm0.03 $ & $0.60\pm0.02 $ & $7.038  \pm0.120 $ & 118/6 \\
                 & $0.240\pm0.002$ & $1.00\pm0.02 $ & $0.20\pm0.03 $ & $4.561  \pm0.240 $ & 71/3 \\
                 & $0.150\pm0.003$ & $1.82\pm0.03 $ & $0.10\pm0.02 $ & $3.306  \pm0.200 $ & 29/2 \\
 \hline
\end{tabular}%
\end{center}}
\end{table*}

\clearpage

\begin{table*}[!htb]
{\scriptsize Table A2. Values of $T$, $y_{\max}$, $y_{\min}$,
$N_0$, $\chi^2$, and ndof corresponding to the curves in Fig. 2 in
which different data are measured in different $\theta$ and $z$
ranges. In the table, $z$ is in the units of cm, and $\theta$ is
not listed, which appears in Fig. 2. In one case, ndof is negative
which appears in terms of ``$-$" and the corresponding curve is
just for eye guiding only.} {\tiny
\begin{center} \vskip-0.3cm
\begin{tabular} {cccccccccccc}\\ \hline\hline Figure &
  $T$ (GeV) & $y_{\max}$& $y_{\min}$& $N_0(\times0.001)$ & $\chi^2$/ndof \\
\hline
                 & $0.320\pm0.005$ & $2.30\pm0.02 $ & $1.10\pm0.02 $ & $0.077  \pm0.002 $ & 33/2 \\
  Fig. 2(a)      & $0.138\pm0.004$ & $3.00\pm0.03 $ & $1.72\pm0.03 $ & $0.702  \pm0.020 $ & 116/12\\
0$\leq$$z$$<$18  & $0.195\pm0.004$ & $2.40\pm0.02 $ & $1.00\pm0.03 $ & $1.987  \pm0.060 $ & 71/12\\
                 & $0.208\pm0.003$ & $2.36\pm0.04 $ & $0.70\pm0.03 $ & $4.859  \pm0.100 $ & 88/9 \\
                 & $0.230\pm0.003$ & $2.10\pm0.02 $ & $0.40\pm0.04 $ & $8.318  \pm0.120 $ & 18/9 \\
 \hline
                 & $0.260\pm0.004$ & $1.80\pm0.02 $ & $0.40\pm0.02 $ & $11.530 \pm0.200 $ & 33/9 \\
  Fig. 2(b)      & $0.169\pm0.002$ & $2.14\pm0.05 $ & $0.65\pm0.03 $ & $13.579 \pm0.200 $ & 18/9 \\
0$\leq$$z$$<$18  & $0.168\pm0.004$ & $2.05\pm0.03 $ & $0.30\pm0.03 $ & $14.905 \pm0.300 $ & 27/6 \\
                 & $0.200\pm0.003$ & $1.60\pm0.02 $ & $0.50\pm0.03 $ & $14.783 \pm0.300 $ & 50/6 \\
                 & $0.175\pm0.003$ & $1.77\pm0.02 $ & $0.35\pm0.04 $ & $15.571 \pm0.300 $ & 37/6 \\
 \hline
                 & $0.210\pm0.030$ & $1.35\pm0.03 $ & $0.25\pm0.02 $ & $15.253 \pm0.200 $ & 39/6 \\
  Fig. 2(c)      & $0.138\pm0.002$ & $2.00\pm0.04 $ & $0.50\pm0.03 $ & $27.457 \pm0.400 $ & 28/3 \\
0$\leq$$z$$<$18  & $0.138\pm0.002$ & $1.75\pm0.04 $ & $0.60\pm0.03 $ & $23.691 \pm0.400 $ & 38/2 \\
                 & $0.155\pm0.003$ & $1.05\pm0.02 $ & $0.65\pm0.02 $ & $21.444 \pm0.400 $ & 12/1 \\
                 & $0.160\pm0.003$ & $0.85\pm0.02 $ & $0.75\pm0.03 $ & $30.222 \pm0.400 $ & 2/$-$\\
\hline
                 & $0.280\pm0.004$ & $2.40\pm0.03 $ & $0.06\pm0.03 $ & $0.313  \pm0.020 $ & 57/3 \\
  Fig. 2(d)      & $0.138\pm0.002$ & $3.00\pm0.02 $ & $1.52\pm0.03 $ & $3.618  \pm0.140 $ & 73/12\\
18$\leq$$z$$<$36 & $0.225\pm0.003$ & $2.30\pm0.02 $ & $1.00\pm0.02 $ & $11.850 \pm0.200 $ & 84/12\\
                 & $0.208\pm0.003$ & $2.36\pm0.03 $ & $0.70\pm0.03 $ & $19.958 \pm0.300 $ & 69/9 \\
                 & $0.228\pm0.002$ & $2.10\pm0.01 $ & $0.50\pm0.02 $ & $24.746 \pm0.400 $ & 9/9 \\
 \hline
                 & $0.262\pm0.002$ & $1.70\pm0.02 $ & $0.20\pm0.02 $ & $25.438 \pm0.300 $ & 57/9 \\
  Fig. 2(e)      & $0.182\pm0.004$ & $2.00\pm0.01 $ & $0.65\pm0.03 $ & $25.160 \pm0.200 $ & 64/9 \\
18$\leq$$z$$<$36 & $0.169\pm0.003$ & $2.05\pm0.03 $ & $0.30\pm0.05 $ & $24.000 \pm0.300 $ & 21/6 \\
                 & $0.200\pm0.005$ & $1.60\pm0.03 $ & $0.40\pm0.02 $ & $22.373 \pm0.300 $ & 30/6 \\
                 & $0.175\pm0.002$ & $1.77\pm0.01 $ & $0.35\pm0.05 $ & $21.727 \pm0.400 $ & 55/6 \\
 \hline
                 & $0.270\pm0.004$ & $1.00\pm0.02 $ & $0.35\pm0.03 $ & $19.006 \pm0.200 $ & 84/6 \\
  Fig. 2(f)      & $0.139\pm0.003$ & $2.00\pm0.04 $ & $0.50\pm0.03 $ & $35.492 \pm0.600 $ & 28/3 \\
18$\leq$$z$$<$36 & $0.138\pm0.002$ & $1.75\pm0.05 $ & $0.60\pm0.03 $ & $31.840 \pm0.600 $ & 52/2 \\
                 & $0.155\pm0.003$ & $1.00\pm0.02 $ & $0.65\pm0.02 $ & $28.264 \pm0.600 $ & 7/1 \\
                 & $0.080\pm0.005$ & $1.40\pm0.02 $ & $1.16\pm0.03 $ & $23.464 \pm0.600 $ & 176/0 \\
\hline
                 & $0.280\pm0.004$ & $2.40\pm0.03 $ & $0.06\pm0.03 $ & $0.690  \pm0.020 $ & 36/3 \\
  Fig. 2(g)      & $0.149\pm0.003$ & $3.00\pm0.03 $ & $1.52\pm0.04 $ & $7.345  \pm0.159 $ & 95/12\\
36$\leq$$z$$<$54 & $0.225\pm0.003$ & $2.45\pm0.03 $ & $0.60\pm0.02 $ & $16.629 \pm0.360 $ & 79/12\\
                 & $0.208\pm0.002$ & $2.36\pm0.03 $ & $0.60\pm0.03 $ & $19.895 \pm0.300 $ & 58/9 \\
                 & $0.218\pm0.002$ & $2.10\pm0.02 $ & $0.30\pm0.03 $ & $20.966 \pm0.400 $ & 33/9 \\
 \hline
                 & $0.260\pm0.003$ & $1.70\pm0.02 $ & $0.20\pm0.02 $ & $22.186 \pm0.400 $ & 48/9 \\
  Fig. 2(h)      & $0.182\pm0.002$ & $2.00\pm0.03 $ & $0.55\pm0.03 $ & $20.936 \pm0.360 $ & 63/9 \\
36$\leq$$z$$<$54 & $0.170\pm0.002$ & $2.05\pm0.03 $ & $0.30\pm0.03 $ & $20.232 \pm0.300 $ & 25/6 \\
                 & $0.210\pm0.004$ & $1.60\pm0.04 $ & $0.40\pm0.03 $ & $17.743 \pm0.200 $ & 9/6 \\
                 & $0.175\pm0.002$ & $1.77\pm0.01 $ & $0.35\pm0.02 $ & $18.830 \pm0.300 $ & 41/6 \\
 \hline
                 & $0.270\pm0.002$ & $1.00\pm0.01 $ & $0.15\pm0.01 $ & $16.232 \pm0.200 $ & 74/6 \\
  Fig. 2(i)      & $0.140\pm0.002$ & $2.00\pm0.01 $ & $0.50\pm0.02 $ & $31.730 \pm0.640 $ & 30/3 \\
36$\leq$$z$$<$54 & $0.138\pm0.002$ & $1.75\pm0.02 $ & $0.40\pm0.02 $ & $24.610 \pm0.400 $ & 47/2 \\
                 & $0.155\pm0.002$ & $1.05\pm0.02 $ & $0.70\pm0.03 $ & $23.472 \pm0.600 $ & 9/1 \\
                 & $0.090\pm0.003$ & $1.40\pm0.02 $ & $0.16\pm0.01 $ & $19.198 \pm0.400 $ & 8/0 \\
\hline
                 & $0.280\pm0.004$ & $2.50\pm0.03 $ & $0.06\pm0.03 $ & $0.943  \pm0.020 $ & 41/3 \\
  Fig. 2(j)      & $0.149\pm0.003$ & $3.05\pm0.02 $ & $1.52\pm0.03 $ & $7.362  \pm0.200 $ & 65/12\\
54$\leq$$z$$<$72 & $0.225\pm0.004$ & $2.40\pm0.02 $ & $0.60\pm0.02 $ & $13.744 \pm0.200 $ & 61/12\\
                 & $0.208\pm0.003$ & $2.36\pm0.01 $ & $0.60\pm0.03 $ & $15.998 \pm0.200 $ & 60/9 \\
                 & $0.220\pm0.003$ & $2.08\pm0.02 $ & $0.30\pm0.03 $ & $16.243 \pm0.240 $ & 13/9 \\
 \hline
                 & $0.260\pm0.004$ & $1.70\pm0.02 $ & $0.30\pm0.02 $ & $16.868 \pm0.240 $ & 50/9 \\
  Fig. 2(k)      & $0.184\pm0.002$ & $2.00\pm0.02 $ & $0.55\pm0.03 $ & $15.986 \pm0.200 $ & 70/9 \\
54$\leq$$z$$<$72 & $0.170\pm0.003$ & $2.05\pm0.03 $ & $0.20\pm0.02 $ & $15.533 \pm0.200 $ & 28/6 \\
                 & $0.210\pm0.004$ & $1.55\pm0.02 $ & $0.20\pm0.03 $ & $13.369 \pm0.200 $ & 48/6 \\
                 & $0.175\pm0.003$ & $1.77\pm0.03 $ & $0.35\pm0.02 $ & $14.123 \pm0.200 $ & 25/6 \\
 \hline
                 & $0.280\pm0.004$ & $1.00\pm0.03 $ & $0.15\pm0.03 $ & $12.808 \pm0.100 $ & 66/6 \\
  Fig. 2(l)      & $0.140\pm0.002$ & $2.00\pm0.02 $ & $0.40\pm0.02 $ & $23.940 \pm0.400 $ & 31/3 \\
54$\leq$$z$$<$72 & $0.138\pm0.002$ & $1.75\pm0.03 $ & $0.40\pm0.04 $ & $19.141 \pm0.400 $ & 44/2 \\
                 & $0.155\pm0.002$ & $1.05\pm0.03 $ & $0.70\pm0.02 $ & $15.698 \pm0.400 $ & 16/1 \\
                 & $0.135\pm0.003$ & $2.70\pm0.02 $ & $0.66\pm0.02 $ & $15.298 \pm0.400 $ & 26/1 \\
\hline
                 & $0.280\pm0.002$ & $2.50\pm0.03 $ & $0.06\pm0.02 $ & $1.019  \pm0.020 $ & 43/3 \\
  Fig. 2(m)      & $0.149\pm0.003$ & $3.05\pm0.03 $ & $1.52\pm0.03 $ & $6.509  \pm0.240 $ & 73/12\\
72$\leq$$z$$<$90 & $0.225\pm0.002$ & $2.37\pm0.02 $ & $0.60\pm0.03 $ & $9.724  \pm0.200 $ & 88/12\\
                 & $0.208\pm0.002$ & $2.30\pm0.03 $ & $0.40\pm0.03 $ & $11.638 \pm0.200 $ & 86/9 \\
                 & $0.220\pm0.002$ & $2.02\pm0.02 $ & $0.25\pm0.02 $ & $12.192 \pm0.200 $ & 23/9 \\
 \hline
                 & $0.260\pm0.003$ & $1.60\pm0.02 $ & $0.30\pm0.02 $ & $12.207 \pm0.200 $ & 51/9 \\
  Fig. 2(n)      & $0.188\pm0.002$ & $1.90\pm0.03 $ & $0.55\pm0.03 $ & $11.864 \pm0.200 $ & 33/9 \\
72$\leq$$z$$<$90 & $0.170\pm0.003$ & $2.05\pm0.02 $ & $0.20\pm0.03 $ & $11.571 \pm0.200 $ & 7/6 \\
                 & $0.210\pm0.003$ & $1.55\pm0.02 $ & $0.20\pm0.02 $ & $11.074 \pm0.200 $ & 45/6 \\
                 & $0.175\pm0.003$ & $1.77\pm0.02 $ & $0.35\pm0.03 $ & $11.769 \pm0.200 $ & 41/6 \\
 \hline
                 & $0.280\pm0.003$ & $1.00\pm0.03 $ & $0.15\pm0.04 $ & $10.322 \pm0.200 $ & 50/6 \\
  Fig. 2(o)      & $0.140\pm0.003$ & $2.00\pm0.04 $ & $0.30\pm0.03 $ & $18.405 \pm0.400 $ & 23/3 \\
72$\leq$$z$$<$90 & $0.138\pm0.004$ & $1.75\pm0.04 $ & $0.40\pm0.05 $ & $15.381 \pm0.400 $ & 30/2 \\
                 & $0.155\pm0.004$ & $1.05\pm0.03 $ & $0.70\pm0.03 $ & $12.841 \pm0.600 $ & 15/1 \\
                 & $0.135\pm0.002$ & $2.70\pm0.04 $ & $0.06\pm0.04 $ & $11.641 \pm0.400 $ & 19/1 \\
\hline
                 & $0.280\pm0.006$ & $2.60\pm0.04 $ & $0.06\pm0.01$  & $9.687  \pm0.200 $ & 47/3 \\
  Fig. 2(p)      & $0.147\pm0.002$ & $3.10\pm0.02 $ & $1.32\pm0.02 $ & $14.156 \pm0.140 $ & 18/12\\
 $z$$=$90        & $0.225\pm0.003$ & $2.32\pm0.03 $ & $0.60\pm0.03 $ & $11.283 \pm0.200 $ & 70/12\\
                 & $0.208\pm0.002$ & $2.30\pm0.02 $ & $0.40\pm0.03 $ & $10.248 \pm0.160 $ & 66/9 \\
                 & $0.220\pm0.003$ & $2.00\pm0.02 $ & $0.30\pm0.02 $ & $8.134  \pm0.200 $ & 64/9 \\
 \hline
                 & $0.260\pm0.004$ & $1.55\pm0.01 $ & $0.30\pm0.02 $ & $12.255 \pm0.400 $ & 91/9 \\
  Fig. 2(q)      & $0.188\pm0.004$ & $1.80\pm0.02 $ & $0.25\pm0.02 $ & $8.593  \pm0.400 $ & 30/6 \\
 $z$$=$90        & $0.165\pm0.002$ & $1.80\pm0.03 $ & $0.20\pm0.03 $ & $5.816  \pm0.320 $ & 37/6 \\
                 & $0.210\pm0.005$ & $1.30\pm0.01 $ & $0.10\pm0.03 $ & $4.394  \pm0.120 $ & 13/3 \\
                 & $0.175\pm0.003$ & $1.33\pm0.03 $ & $0.05\pm0.04 $ & $3.019  \pm0.040 $ & 34/2 \\
 \hline
\end{tabular}%
\end{center}}
\end{table*}

\begin{table*}[!htb]
{\scriptsize Table A3. Values of $T$, $y_{\max}$, $y_{\min}$,
$N_0$, $\chi^2$, and ndof corresponding to the curves in Fig. 3 in
which different data are measured in different $z$ and $\theta$
ranges. In the table, $\theta$ is in the units of mrad, and $z$ is
not listed, which appears in Fig. 3.} {\tiny
\begin{center} \vskip-0.3cm
\begin{tabular} {cccccccccccc}\\ \hline\hline Figure &
  $T$ (GeV) & $y_{\max}$& $y_{\min}$& $N_0(\times0.001)$ & $\chi^2$/ndof \\
\hline
                        & $0.400\pm0.004$ & $1.91\pm0.02 $ & $1.50\pm0.02 $ & $1.622 \pm0.040 $ & 48/12\\
                        & $0.400\pm0.004$ & $1.92\pm0.03 $ & $1.37\pm0.02 $ & $6.272 \pm0.100 $ & 57/12\\
  Fig. 3(a)             & $0.400\pm0.004$ & $1.97\pm0.02 $ & $1.37\pm0.02 $ & $10.672\pm0.200 $ & 95/12\\
20$\leq$$\theta$$<$40   & $0.400\pm0.003$ & $2.05\pm0.01 $ & $1.30\pm0.02 $ & $11.335\pm0.240 $ & 55/12\\
                        & $0.400\pm0.005$ & $2.05\pm0.02 $ & $1.30\pm0.02 $ & $8.904 \pm0.240 $ & 68/12\\
                        & $0.340\pm0.001$ & $2.22\pm0.01 $ & $1.20\pm0.02 $ & $19.874\pm0.400 $ & 231/12\\
 \hline
                        & $0.200\pm0.004$ & $2.01\pm0.03 $ & $0.80\pm0.02 $ & $30.276\pm0.400 $ & 30/9 \\
                        & $0.230\pm0.003$ & $1.90\pm0.02 $ & $0.20\pm0.02 $ & $58.805\pm1.600 $ & 19/9 \\
  Fig. 3(b)             & $0.228\pm0.002$ & $1.89\pm0.03 $ & $0.20\pm0.02 $ & $46.620\pm1.600 $ & 11/9 \\
100$\leq$$\theta$$<$140 & $0.230\pm0.003$ & $1.86\pm0.02 $ & $0.10\pm0.02 $ & $34.882\pm0.800 $ & 17/9 \\
                        & $0.220\pm0.004$ & $1.85\pm0.03 $ & $0.35\pm0.02 $ & $26.181\pm0.600 $ & 20/9 \\
                        & $0.235\pm0.003$ & $1.75\pm0.02 $ & $0.10\pm0.02 $ & $12.809\pm0.400 $ & 37/9 \\
 \hline
                        & $0.440\pm0.005$ & $1.80\pm0.02 $ & $0.30\pm0.03 $ & $1.034 \pm0.020 $ & 57/12\\
                        & $0.400\pm0.004$ & $1.92\pm0.03 $ & $0.50\pm0.03 $ & $4.136 \pm0.200 $ & 16/12\\
  Fig. 3(c)             & $0.400\pm0.004$ & $1.97\pm0.03 $ & $0.30\pm0.02 $ & $7.232 \pm0.160 $ & 52/12\\
20$\leq$$\theta$$<$40   & $0.400\pm0.005$ & $2.00\pm0.03 $ & $0.30\pm0.03 $ & $7.329 \pm0.200 $ & 23/12\\
                        & $0.444\pm0.004$ & $1.82\pm0.02 $ & $0.50\pm0.02 $ & $5.935 \pm0.100 $ & 26/12\\
                        & $0.345\pm0.005$ & $2.08\pm0.02 $ & $0.60\pm0.02 $ & $12.610\pm0.200 $ & 40/12\\
\hline
                        & $0.202\pm0.002$ & $2.03\pm0.02 $ & $0.30\pm0.02 $ & $24.814\pm0.480 $ & 17/9 \\
                        & $0.230\pm0.005$ & $1.86\pm0.02 $ & $0.10\pm0.02 $ & $48.532\pm0.800 $ & 15/9 \\
  Fig. 3(d)             & $0.230\pm0.004$ & $1.81\pm0.02 $ & $0.10\pm0.02 $ & $40.351\pm0.600 $ & 36/9 \\
100$\leq$$\theta$$<$140 & $0.240\pm0.004$ & $1.73\pm0.02 $ & $0.23\pm0.02 $ & $30.234\pm0.800 $ & 11/9 \\
                        & $0.245\pm0.002$ & $1.70\pm0.02 $ & $0.22\pm0.02 $ & $23.424\pm0.520 $ & 27/9 \\
                        & $0.245\pm0.002$ & $1.68\pm0.02 $ & $0.10\pm0.02 $ & $10.480\pm0.520 $ & 36/9 \\
 \hline
\end{tabular}%
\end{center}}
\end{table*}

\begin{table*}[!htb]
{\scriptsize Table A4. Values of $T$, $y_{\max}$, $y_{\min}$,
$N_0$, $\chi^2$, and ndof corresponding to the curves in Fig. 4 in
which different data are measured in different $\theta$ and $z$
ranges. In the table, $z$ is in the units of cm, and $\theta$ is
not listed, which appears in Fig. 4.} {\tiny
\begin{center} \vskip-0.3cm
\begin{tabular} {cccccccccccc}\\ \hline\hline Figure &
  $T$ (GeV) & $y_{\max}$ & $y_{\min}$& $N_0(\times0.001)$ & $\chi^2$/ndof \\
\hline
                 & $0.300\pm0.003$ & $2.00\pm0.03 $ & $1.20\pm0.03 $ & $0.335 \pm0.012 $ & 18/2 \\
  Fig. 4(a)      & $0.300\pm0.003$ & $1.90\pm0.02 $ & $1.10\pm0.03 $ & $3.079 \pm0.120 $ & 30/2 \\
0$\leq$$z$$<$18  & $0.300\pm0.003$ & $1.45\pm0.03 $ & $1.10\pm0.03 $ & $4.787 \pm0.180 $ & 24/2 \\
                 & $0.300\pm0.004$ & $1.10\pm0.02 $ & $0.60\pm0.03 $ & $6.462 \pm0.300 $ & 31/2 \\
 \hline
                 & $0.400\pm0.003$ & $2.00\pm0.03 $ & $1.20\pm0.03 $ & $2.569 \pm0.060 $ & 8/2 \\
  Fig. 4(b)      & $0.300\pm0.003$ & $2.00\pm0.03 $ & $1.10\pm0.02 $ & $9.552 \pm0.300 $ & 26/2 \\
18$\leq$$z$$<$36 & $0.400\pm0.004$ & $1.00\pm0.02 $ & $0.95\pm0.01 $ & $8.080 \pm0.180 $ & 27/2 \\
                 & $0.320\pm0.004$ & $1.08\pm0.02 $ & $0.70\pm0.03 $ & $8.773 \pm0.100 $ & 25/2 \\
 \hline
                 & $0.420\pm0.004$ & $1.75\pm0.02 $ & $1.35\pm0.02 $ & $3.828 \pm0.060 $ & 47/4 \\
  Fig. 4(c)      & $0.300\pm0.004$ & $1.85\pm0.04 $ & $1.10\pm0.03 $ & $7.922 \pm0.180 $ & 39/2 \\
36$\leq$$z$$<$54 & $0.280\pm0.004$ & $1.45\pm0.04 $ & $1.35\pm0.03 $ & $6.561 \pm0.240 $ & 21/2 \\
                 & $0.316\pm0.004$ & $1.15\pm0.03 $ & $0.40\pm0.04 $ & $7.016 \pm0.500 $ & 14/2 \\
\hline
                 & $0.400\pm0.005$ & $2.00\pm0.03 $ & $1.30\pm0.02 $ & $4.065 \pm0.180 $ & 38/4 \\
  Fig. 4(d)      & $0.300\pm0.003$ & $1.80\pm0.03 $ & $1.05\pm0.03 $ & $5.934 \pm0.300 $ & 18/2 \\
54$\leq$$z$$<$72 & $0.360\pm0.005$ & $1.35\pm0.03 $ & $0.60\pm0.06 $ & $4.578 \pm0.180 $ & 13/2 \\
                 & $0.275\pm0.002$ & $1.02\pm0.01 $ & $1.00\pm0.01 $ & $5.361 \pm0.200 $ & 38/2 \\
 \hline
                 & $0.400\pm0.004$ & $2.20\pm0.03 $ & $1.35\pm0.03 $ & $3.806 \pm0.120 $ & 14/4 \\
  Fig. 4(e)      & $0.300\pm0.004$ & $1.80\pm0.03 $ & $1.05\pm0.03 $ & $4.641 \pm0.120 $ & 32/2 \\
72$\leq$$z$$<$90 & $0.280\pm0.003$ & $1.45\pm0.03 $ & $1.20\pm0.03 $ & $3.491 \pm0.120 $ & 17/2 \\
                 & $0.300\pm0.004$ & $1.10\pm0.03 $ & $0.50\pm0.04 $ & $4.315 \pm0.150 $ & 22/2 \\
 \hline
  Fig. 4(f)      & $0.400\pm0.003$ & $2.20\pm0.03 $ & $1.00\pm0.03 $ & $6.300 \pm0.120 $ & 10/5 \\
 $z$$=$90        & $0.300\pm0.006$ & $1.79\pm0.03 $ & $1.00\pm0.02 $ & $3.123 \pm0.060 $ & 36/2 \\
\hline
\end{tabular}%
\end{center}}
\end{table*}

\begin{table*}[!htb]
{\scriptsize Table A5. Values of $T$, $y_{\max}$, $y_{\min}$,
$N_0$, $\chi^2$, and ndof corresponding to the curves in Fig. 5 in
which different data are measured in different $\theta$ and $z$
ranges. In the table, $z$ is in the units of cm, and $\theta$ is
not listed, which appears in Fig. 5.} {\tiny
\begin{center} \vskip-0.3cm
\begin{tabular} {cccccccccccc}\\ \hline\hline Figure &
  $T$ (GeV) & $y_{\max}$& $y_{\min}$& $N_0(\times 0.001)$ & $\chi^2$/ndof \\
\hline
                 & $0.280\pm0.005$ & $1.90\pm0.03 $ & $1.80\pm0.03 $ & $0.191 \pm0.006 $ & 27/2 \\
  Fig. 5(a)      & $0.300\pm0.005$ & $1.70\pm0.03 $ & $1.10\pm0.02 $ & $1.173 \pm0.060 $ & 18/2 \\
0$\leq$$z$$<$18  & $0.300\pm0.003$ & $1.20\pm0.01 $ & $1.10\pm0.01 $ & $1.997 \pm0.060 $ & 67/2 \\
                 & $0.290\pm0.005$ & $1.10\pm0.04 $ & $0.50\pm0.02 $ & $2.685 \pm0.100 $ & 19/2 \\
 \hline
                 & $0.400\pm0.005$ & $1.73\pm0.03 $ & $1.20\pm0.03 $ & $0.966 \pm0.060 $ & 39/2 \\
  Fig. 5(b)      & $0.300\pm0.003$ & $1.50\pm0.03 $ & $1.30\pm0.03 $ & $3.386 \pm0.120 $ & 72/2 \\
18$\leq$$z$$<$36 & $0.400\pm0.003$ & $1.00\pm0.02 $ & $0.70\pm0.01 $ & $2.796 \pm0.090 $ & 18/2 \\
                 & $0.320\pm0.003$ & $0.91\pm0.02 $ & $0.70\pm0.02 $ & $3.752 \pm0.150 $ & 25/2 \\
 \hline
                 & $0.420\pm0.005$ & $1.45\pm0.02 $ & $1.40\pm0.01 $ & $1.247 \pm0.030 $ & 33/4 \\
  Fig. 5(c)      & $0.340\pm0.003$ & $1.65\pm0.03 $ & $1.10\pm0.03 $ & $3.308 \pm0.150 $ & 18/2 \\
36$\leq$$z$$<$54 & $0.248\pm0.003$ & $1.45\pm0.03 $ & $1.30\pm0.03 $ & $2.580 \pm0.090 $ & 18/2 \\
                 & $0.300\pm0.003$ & $1.10\pm0.03 $ & $0.60\pm0.01 $ & $2.800 \pm0.100 $ & 29/2 \\
\hline
                 & $0.440\pm0.010$ & $1.45\pm0.07 $ & $1.30\pm0.05 $ & $1.211 \pm0.060 $ & 67/4 \\
  Fig. 5(d)      & $0.300\pm0.004$ & $1.70\pm0.03 $ & $1.05\pm0.03 $ & $2.225 \pm0.060 $ & 21/2 \\
54$\leq$$z$$<$72 & $0.330\pm0.005$ & $1.10\pm0.03 $ & $1.03\pm0.03 $ & $2.058 \pm0.090 $ & 17/2 \\
                 & $0.288\pm0.002$ & $1.00\pm0.02 $ & $0.95\pm0.02 $ & $2.131 \pm0.100 $ & 55/2 \\
 \hline
                 & $0.330\pm0.004$ & $1.85\pm0.03 $ & $1.28\pm0.02 $ & $0.894 \pm0.060 $ & 39/4 \\
  Fig. 5(e)      & $0.260\pm0.004$ & $2.00\pm0.04 $ & $0.60\pm0.04 $ & $1.811 \pm0.060 $ & 10/2 \\
72$\leq$$z$$<$90 & $0.245\pm0.004$ & $1.45\pm0.03 $ & $1.00\pm0.02 $ & $1.496 \pm0.060 $ & 5/2 \\
                 & $0.300\pm0.004$ & $0.95\pm0.03 $ & $0.63\pm0.02 $ & $1.773 \pm0.100 $ & 23/2 \\
 \hline
  Fig. 5(f)      & $0.350\pm0.003$ & $2.00\pm0.02 $ & $1.00\pm0.02 $ & $1.640 \pm0.060 $ & 36/5 \\
 $z$$=$90        & $0.300\pm0.004$ & $1.60\pm0.04 $ & $1.20\pm0.03 $ & $1.283 \pm0.060 $ & 39/2 \\
\hline
\end{tabular}%
\end{center}}
\end{table*}

\clearpage

\begin{table*}[!htb]
{\scriptsize Table A6. Values of $T$, $k$, $y_{\max}$, $y_{\min}$,
$y_{L\max}$, $y_{L\min}$, $N_0$, $\chi^2$, and ndof corresponding
to the curves in Fig. 6 in which different data are measured in
different $\theta$ and $z$ ranges. In the table, $z$ is in the
units of cm, and $\theta$ is not listed, which appears in Fig. 6.
In a few cases, ndof are negative which appear in terms of ``$-$"
and the corresponding curves are just for eye guiding only.
\begin{center} \vskip-0.3cm
\begin{tabular} {cccccccccccc}\\ \hline\hline Figure &
$T$ (GeV) &$k$ & $y_{\max}$& $y_{\min}$& $y_{L\max}$& $y_{L\min}$& $N_0(\times 0.001)$ & $\chi^2$/ndof \\
\hline
                 & $0.400\pm0.004$ & $0.40\pm0.01$ & $1.60\pm0.05$ & $1.00\pm0.10$ & $4.00\pm0.07$ & $3.55\pm0.08$ & $2.996\pm0.200$  & 10/$-$\\
  Fig. 6(a)      & $0.280\pm0.004$ & $0.40\pm0.02$ & $1.50\pm0.10$ & $1.20\pm0.10$ & $3.10\pm0.10$ & $2.70\pm0.10$ & $1.337\pm0.020$  & 118/9\\
0$\leq$$z$$<$18  & $0.200\pm0.005$ & $0.40\pm0.02$ & $1.50\pm0.08$ & $1.20\pm0.10$ & $3.45\pm0.10$ & $2.00\pm0.08$ & $2.145\pm0.100$  & 86/7\\
                 & $0.127\pm0.003$ & $0.40\pm0.01$ & $1.55\pm0.07$ & $1.50\pm0.06$ & $3.45\pm0.05$ & $2.20\pm0.05$ & $7.216\pm0.160$  & 202/6\\
                 & $0.125\pm0.005$ & $0.45\pm0.01$ & $1.55\pm0.05$ & $0.90\pm0.05$ & $3.45\pm0.08$ & $2.00\pm0.05$ & $10.156\pm0.400$ & 128/6\\
\hline
                 & $0.115\pm0.003$ & $0.40\pm0.01$ & $0.98\pm0.02$ & $0.80\pm0.02$ & $3.90\pm0.08$ & $1.30\pm0.05$ & $11.629\pm0.400$ & 50/3\\
  Fig. 6(b)      & $0.150\pm0.004$ & $0.54\pm0.02$ & $0.80\pm0.03$ & $0.70\pm0.03$ & $2.80\pm0.07$ & $0.60\pm0.10$ & $13.653\pm0.400$ & 57/3\\
0$\leq$$z$$<$18  & $0.164\pm0.004$ & $0.53\pm0.02$ & $0.80\pm0.03$ & $0.70\pm0.03$ & $3.50\pm0.10$ & $0.60\pm0.20$ & $13.653\pm0.400$ & 57/$-$\\
                 & $0.177\pm0.005$ & $0.53\pm0.02$ & $0.80\pm0.05$ & $0.70\pm0.05$ & $3.00\pm0.12$ & $1.50\pm0.10$ & $12.291\pm0.800$ & 28/$-$\\
                 & $0.140\pm0.010$ & $0.35\pm0.02$ & $0.40\pm0.05$ & $0.30\pm0.05$ & $3.50\pm0.10$ & $1.00\pm0.05$ & $27.176\pm0.800$ & 5/$-$\\
\hline
                 & $0.450\pm0.010$ & $0.53\pm0.02$ & $1.60\pm0.10$ & $1.00\pm0.10$ & $4.00\pm0.10$ & $3.55\pm0.05$ & $8.988\pm0.400$  & 9/$-$\\
  Fig. 6(c)      & $0.280\pm0.006$ & $0.43\pm0.02$ & $1.50\pm0.20$ & $1.20\pm0.20$ & $3.10\pm0.03$ & $2.70\pm0.03$ & $7.496\pm0.300$  & 117/9\\
18$\leq$$z$$<$36 & $0.200\pm0.003$ & $0.49\pm0.01$ & $1.50\pm0.05$ & $1.20\pm0.06$ & $3.45\pm0.05$ & $2.10\pm0.03$ & $14.518\pm0.400$ & 42/7\\
                 & $0.130\pm0.006$ & $0.45\pm0.02$ & $1.60\pm0.02$ & $1.30\pm0.03$ & $3.45\pm0.03$ & $2.20\pm0.01$ & $28.178\pm0.800$ & 171/6\\
                 & $0.125\pm0.005$ & $0.45\pm0.02$ & $1.55\pm0.03$ & $0.85\pm0.03$ & $3.45\pm0.05$ & $2.00\pm0.05$ & $22.773\pm0.480$ & 118/7\\
\hline
                 & $0.115\pm0.003$ & $0.40\pm0.01$ & $1.00\pm0.03$ & $0.80\pm0.03$ & $3.90\pm0.04$ & $1.40\pm0.04$ & $20.745\pm0.600$ & 40/3\\
  Fig. 6(d)      & $0.120\pm0.003$ & $0.54\pm0.01$ & $0.65\pm0.03$ & $0.60\pm0.03$ & $2.80\pm0.05$ & $0.90\pm0.05$ & $20.636\pm0.600$ & 50/3\\
18$\leq$$z$$<$36 & $0.158\pm0.003$ & $0.57\pm0.03$ & $0.60\pm0.04$ & $0.50\pm0.04$ & $3.50\pm0.10$ & $0.80\pm0.05$ & $19.404\pm0.400$ & 18/$-$\\
                 & $0.167\pm0.002$ & $0.55\pm0.02$ & $0.65\pm0.02$ & $0.55\pm0.02$ & $3.00\pm0.30$ & $1.50\pm0.08$ & $19.730\pm0.400$ & 18/$-$\\
                 & $0.180\pm0.006$ & $0.34\pm0.01$ & $0.40\pm0.03$ & $0.36\pm0.02$ & $3.50\pm0.06$ & $1.50\pm0.02$ & $47.883\pm0.800$ & 6/$-$\\
\hline
                 & $0.455\pm0.005$ & $0.53\pm0.01$ & $1.60\pm0.05$ & $1.00\pm0.06$ & $4.06\pm0.06$ & $3.55\pm0.05$ & $18.555\pm0.400$ & 11/$-$\\
  Fig. 6(e)      & $0.290\pm0.004$ & $0.53\pm0.01$ & $1.50\pm0.15$ & $1.20\pm0.15$ & $3.10\pm0.02$ & $2.70\pm0.03$ & $15.862\pm0.300$ & 94/9\\
36$\leq$$z$$<$54 & $0.200\pm0.005$ & $0.50\pm0.01$ & $1.50\pm0.08$ & $1.20\pm0.10$ & $3.45\pm0.05$ & $2.10\pm0.03$ & $22.391\pm0.600$ & 35/7\\
                 & $0.130\pm0.004$ & $0.45\pm0.01$ & $1.60\pm0.01$ & $1.30\pm0.03$ & $3.45\pm0.03$ & $2.15\pm0.01$ & $28.281\pm0.400$ & 173/6\\
                 & $0.125\pm0.003$ & $0.45\pm0.02$ & $1.55\pm0.02$ & $0.88\pm0.02$ & $3.45\pm0.02$ & $2.00\pm0.02$ & $20.981\pm0.400$ & 118/6\\
\hline
                 & $0.115\pm0.002$ & $0.35\pm0.01$ & $1.00\pm0.02$ & $0.80\pm0.02$ & $4.00\pm0.03$ & $1.50\pm0.02$ & $19.587\pm0.400$ & 42/3\\
  Fig. 6(f)      & $0.120\pm0.004$ & $0.58\pm0.02$ & $0.70\pm0.02$ & $0.50\pm0.02$ & $2.80\pm0.02$ & $0.90\pm0.03$ & $18.925\pm0.400$ & 48/3\\
36$\leq$$z$$<$54 & $0.164\pm0.001$ & $0.60\pm0.02$ & $0.60\pm0.03$ & $0.50\pm0.03$ & $3.50\pm0.20$ & $0.80\pm0.08$ & $17.285\pm0.400$ & 23/$-$\\
                 & $0.155\pm0.005$ & $0.55\pm0.01$ & $0.65\pm0.04$ & $0.55\pm0.04$ & $3.00\pm0.20$ & $1.50\pm0.10$ & $16.519\pm0.400$ & 12/$-$\\
                 & $0.178\pm0.004$ & $0.43\pm0.01$ & $0.53\pm0.03$ & $0.36\pm0.03$ & $3.50\pm0.15$ & $1.50\pm0.02$ & $37.825\pm0.960$ & 7/$-$\\
\hline
                 & $0.450\pm0.004$ & $0.53\pm0.01$ & $1.60\pm0.20$ & $1.00\pm0.20$ & $4.00\pm0.05$ & $3.55\pm0.05$ & $24.718\pm0.500$ & 13/$-$\\
  Fig. 6(g)      & $0.290\pm0.003$ & $0.53\pm0.01$ & $1.50\pm0.03$ & $1.20\pm0.15$ & $3.10\pm0.02$ & $2.70\pm0.03$ & $19.202\pm0.240$ & 32/9\\
54$\leq$$z$$<$72 & $0.200\pm0.005$ & $0.50\pm0.01$ & $1.50\pm0.10$ & $1.20\pm0.10$ & $3.30\pm0.06$ & $2.10\pm0.05$ & $18.359\pm0.200$ & 31/7\\
                 & $0.130\pm0.002$ & $0.45\pm0.01$ & $1.60\pm0.01$ & $1.30\pm0.01$ & $3.45\pm0.03$ & $2.15\pm0.01$ & $23.629\pm0.400$ & 174/6\\
                 & $0.130\pm0.003$ & $0.45\pm0.01$ & $1.55\pm0.02$ & $0.88\pm0.02$ & $3.45\pm0.03$ & $2.00\pm0.03$ & $17.502\pm0.400$ & 129/6\\
\hline
                 & $0.145\pm0.003$ & $0.45\pm0.02$ & $0.90\pm0.02$ & $0.70\pm0.03$ & $4.00\pm0.10$ & $0.95\pm0.03$ & $15.343\pm0.400$ & 32/3\\
  Fig. 6(h)      & $0.115\pm0.005$ & $0.55\pm0.01$ & $0.65\pm0.03$ & $0.50\pm0.03$ & $3.00\pm0.05$ & $1.00\pm0.05$ & $14.933\pm0.400$ & 66/3\\
54$\leq$$z$$<$72 & $0.160\pm0.003$ & $0.60\pm0.02$ & $0.55\pm0.03$ & $0.50\pm0.03$ & $3.50\pm0.20$ & $0.80\pm0.02$ & $14.798\pm0.400$ & 9/$-$\\
                 & $0.150\pm0.004$ & $0.55\pm0.02$ & $0.65\pm0.05$ & $0.55\pm0.05$ & $3.00\pm0.10$ & $1.50\pm0.05$ & $13.657\pm0.400$ & 16/$-$\\
                 & $0.170\pm0.003$ & $0.43\pm0.01$ & $0.53\pm0.03$ & $0.36\pm0.03$ & $3.50\pm0.07$ & $1.50\pm0.03$ & $29.279\pm0.960$ & 37/$-$\\
\hline
                 & $0.450\pm0.004$ & $0.53\pm0.01$ & $1.60\pm0.03$ & $1.00\pm0.05$ & $4.00\pm0.10$ & $3.55\pm0.10$ & $26.965\pm0.600$ & 6/$-$\\
  Fig. 6(i)      & $0.290\pm0.003$ & $0.53\pm0.01$ & $1.50\pm0.10$ & $1.20\pm0.15$ & $3.10\pm0.03$ & $2.70\pm0.03$ & $17.031\pm0.280$ & 33/9\\
72$\leq$$z$$<$90 & $0.160\pm0.004$ & $0.50\pm0.01$ & $1.50\pm0.03$ & $1.20\pm0.02$ & $3.45\pm0.03$ & $2.10\pm0.03$ & $13.722\pm0.240$ & 19/7\\
                 & $0.140\pm0.004$ & $0.60\pm0.01$ & $1.60\pm0.03$ & $0.50\pm0.03$ & $3.60\pm0.02$ & $1.80\pm0.03$ & $18.999\pm0.480$ & 99/6\\
                 & $0.130\pm0.004$ & $0.50\pm0.01$ & $1.55\pm0.03$ & $0.88\pm0.02$ & $3.45\pm0.03$ & $0.90\pm0.03$ & $13.342\pm0.520$ & 133/6\\
\hline
                 & $0.145\pm0.005$ & $0.53\pm0.01$ & $0.90\pm0.04$ & $0.40\pm0.05$ & $4.00\pm0.04$ & $0.95\pm0.05$ & $12.922\pm0.520$ & 8/3\\
  Fig. 6(j)      & $0.120\pm0.001$ & $0.55\pm0.01$ & $0.65\pm0.02$ & $0.50\pm0.01$ & $3.00\pm0.02$ & $1.00\pm0.03$ & $11.877\pm0.480$ & 51/3\\
72$\leq$$z$$<$90 & $0.160\pm0.002$ & $0.63\pm0.02$ & $0.55\pm0.01$ & $0.50\pm0.01$ & $3.50\pm0.15$ & $0.80\pm0.05$ & $11.686\pm0.320$ & 10/$-$\\
                 & $0.155\pm0.006$ & $0.60\pm0.03$ & $0.65\pm0.03$ & $0.55\pm0.03$ & $3.00\pm0.15$ & $1.20\pm0.06$ & $10.773\pm0.400$ & 25/$-$\\
                 & $0.170\pm0.003$ & $0.50\pm0.01$ & $0.53\pm0.02$ & $0.36\pm0.03$ & $3.50\pm0.20$ & $1.50\pm0.05$ & $20.968\pm0.400$ & 56/$-$\\
\hline
                 & $0.450\pm0.010$ & $0.53\pm0.02$ & $1.60\pm0.10$ & $1.00\pm0.04$ & $4.15\pm0.15$ & $3.55\pm0.15$ & $193.715\pm0.460$& 44/1\\
  Fig. 6(k)      & $0.290\pm0.005$ & $0.53\pm0.01$ & $1.50\pm0.10$ & $1.20\pm0.10$ & $3.10\pm0.01$ & $2.70\pm0.03$ & $36.734\pm0.460$ & 186/9\\
 $z$$=$90        & $0.160\pm0.002$ & $0.55\pm0.01$ & $1.50\pm0.03$ & $1.20\pm0.02$ & $3.45\pm0.04$ & $2.10\pm0.03$ & $15.771\pm0.500$ & 192/9\\
                 & $0.140\pm0.004$ & $0.60\pm0.01$ & $1.60\pm0.03$ & $0.50\pm0.05$ & $3.60\pm0.07$ & $1.80\pm0.06$ & $14.012\pm0.400$ & 104/6\\
\hline
  Fig. 6(l)      & $0.130\pm0.005$ & $0.48\pm0.01$ & $1.55\pm0.03$ & $0.88\pm0.02$ & $3.45\pm0.12$ & $1.90\pm0.03$ & $6.526\pm0.200$  & 146/6\\
 $z$$=$90        & $0.160\pm0.004$ & $0.60\pm0.01$ & $0.65\pm0.05$ & $0.30\pm0.03$ & $2.30\pm0.02$ & $0.60\pm0.04$ & $4.851\pm0.200$  & 30/1\\
                 & $0.120\pm0.001$ & $0.55\pm0.01$ & $0.65\pm0.03$ & $0.50\pm0.02$ & $3.00\pm0.02$ & $1.00\pm0.04$ & $3.167\pm0.200$  & 51/1\\
\hline
\end{tabular}%
\end{center}}
\end{table*}

\end{multicols}
\end{document}